\documentclass{aa}
\usepackage{graphicx}
\usepackage{txfonts}
\usepackage{epstopdf}
\usepackage{microtype}
\usepackage{caption}
\usepackage{array}
\usepackage{subcaption}
\usepackage{float}
\bibpunct{(}{)}{;}{a}{}{,} 

\begin{document}

\title{The radio evolution of supernova SN\,2008iz in M\,82}
\author{N.~ Kimani\inst{\ref{inst1}}\and K.~ Sendlinger\inst{\ref{inst1},\ref{inst2}}\and A.~ Brunthaler\inst{\ref{inst1}}\and K.~ M.~ Menten\inst{\ref{inst1}}\and I.~ Mart\'{i}-vidal \inst{\ref{inst3}}\and C.~ Henkel\inst{\ref{inst1},\ref{inst4}} \and H. Falcke\inst{\ref{inst5},\ref{inst6}}  \and \\T.~W.~B.~ Muxlow\inst{\ref{inst7}} \and  R.~J.~ Beswick\inst{\ref{inst7}} \and G.~C. ~Bower\inst{\ref{inst8}}}

\institute{Max-Planck-Institut f\"{u}r Radioastronomie, Auf dem H\"{u}gel 69, D-53121 Bonn, Germany \label{inst1} \\ nkimani@mpifr-bonn.mpg.de 
\and Argelander-Institut f\"{u}r Astronomie, Auf dem H\"{u}gel 71, D-53121 Bonn, Germany\label{inst2}
\and Onsala Space Observatory, Chalmers Univ. of Technology, 43992, Onsala, Sweden \label{inst3}
\and Astron. Dept., King Abdulaziz University, P.O. Box 80203, Jeddah 21589, Saudi Arabia \label{inst4}
\and Department of Astrophysics, Radboud University Nijmegen, Postbus 9010, NL-6500 GL Nijmegen, The Netherlands \label{inst5}
\and ASTRON, Postbus 2, NL-7990 AA Dwingeloo, The Netherlands \label{inst6}
\and Jodrell Bank Centre for Astrophysics, School of Physics and Astronomy, The university of Manchester, Oxford Road, Manchester M13 9PL, United Kingdom \label{inst7}
\and Academia Sinica Institute of Astronomy and Astrophysics, 645 N. A\'ohoku Place, Hilo, HI 96720, USA \label{inst8}
}


\abstract{We report on multi-frequency Very Large Array (VLA) and Very Long Baseline Interferometry (VLBI) radio observations for a monitoring campaign of supernova SN\,2008iz in the nearby irregular galaxy M$\,$82. We fit two models to the data, a simple time power-law, S$\propto$ t$^\beta$, and a simplified Weiler model, yielding decline indices, $\beta$ of -1.22$\pm$0.07 (days 100-1500) and -1.41$\pm$0.02 (days 76-2167), respectively. The late-time radio light curve evolution shows flux-density flares at $\sim$970 and $\sim$\,1400 days which are a factor of $\sim$\,2 and $\sim$\,4 higher than the expected flux, respectively. The later flare, besides being brighter, does not show signs of decline at least from results examined so far (2014 January 23; day 2167). We derive the spectral index, $\alpha$, S$\propto$ $\nu^\alpha$ for frequencies 1.4 to 43\,GHz for SN\,2008iz during the period from $\sim$\,430 to 2167 days after the supernova explosion. The value of $\alpha$ shows no signs of evolution and remains steep $\approx$$-$1 throughout the period, unlike that of SN\,1993J which started flattening at $\sim$day 970. From the 4.8 and 8.4\,GHz VLBI images, the supernova expansion is seen to start with shell like structure that gets more and more asymmetric, then breaks up in the later epochs with bright structures dominating the southern part of the ring. This structural evolution differs significantly from SN\,1993J, which remains circularly symmetric over 4000 days after the explosion. The VLBI 4.8 and 8.4\,GHz images are used to derive a deceleration index, $m$, for SN\,2008iz, of 0.86\,$\pm$\,0.02, and the average expansion velocity between  days 73 and 1400 as (12.1$\pm$0.2)$\times$10$^{3}$\,km\,s$^{-1}$. From the energy equipartition between magnetic field and particles, we estimate the minimum total energy in relativistic particles and the magnetic fields and particles, we estimate the minimum total energy in relativistic particles and the magnetic fields during the supernova expansion and also find the magnetic field amplification factor for SN\,2008iz to be in the range of 55 -- 400.
} 


\maketitle 

\section{Introduction.} 
Radio-loud supernovae are rare events with just a few dozen detected \citep{Weiler2002}. The majority of them are relatively distant or fairly weak, making them difficult to study in great detail.  To
date, the best known example is SN$\,$1993J in M$\,$81 (\citealp{Martividal2011a,Martividal2011b}
and references therein), which has been studied extensively due to
its proximity (3.63$\,$Mpc, \citealp{Freedmann1994}), environment
(which allows for multi-wavelength studies) and galaxy orientation
(M$\,$81 is almost face-on). The discovery of SN$\,$2008iz \citep{Brunthaler2009a}
offers the possibility to study another supernova at a very similar
distance in great detail and to make a comparison to SN$\,$1993J.
For instance, considering the peak 5$\,$GHz radio luminosities, SN$\,$2008iz
at L$_{5GHz}$$\propto$24.4$\times$10$^{26}$$\,$ergs$^{-1}$Hz$^{-1}$
is comparable to SN$\,$1993J at L$_{5GHz}$$\propto$15.5$\times$10$^{26}$$\,$ergs$^{-1}$Hz$^{-1}$ \citep{VanDyk1994}. The two supernovae are also comparable in the time they took to reach their peak after the explosion, with SN$\,$2008iz
taking $\approx$125 days while SN$\,$1993J took 133$\,$days \citep{Weiler2007}. 
\\
SN$\,$2008iz was discovered in M$\,$82, which is a nearby irregular (IO)
galaxy forming part of M$\,$81 group. This galaxy harbours numerous bright
radio supernova remnants with the current estimate at 50 supernovae
\citep{Muxlow1994,Beswick2006,Fenech2008,Kronberg2000,Weiler2002}. 
\\
Supernova 2008iz was discovered in 2009 April as a bright radio transient
\citep{Brunthaler2009a,Brunthaler2009b}. The explosion date of the
supernova is estimated to be 2008 February 18$\pm$6$\:$days \citep{Brunthaler2009b,Marchili2010}.
The discovery was made at radio wavelengths with the VLA at 22$\,$GHz
\citep{Brunthaler2009a} and confirmed with MERLIN \citep{Muxlow2009,Beswick2009}
and the Urumqi 25 meter telescope at 5$\,$GHz \citep{Marchili2010}. Endeavours to make detections in other astronomical windows have never been successful. For instance, there are no detections in visible light and the X-Ray regimes \citep{Brunthaler2009b}. \citet{Varenius2015} also report the LOFAR (Low-Frequency Array for Radio Astronomy) non-detection at 154$\,$MHz. However, a detection of the near-IR counterpart for SN\,2008iz at $\sim$480 days was reported by \cite{Mattila2013} after careful analyses using image subtraction technique and considering possible uncertainties caused by the differing filter response functions between the Gemini-North telescope and the HST. Overall, this makes it challenging to classify this supernova. Nevertheless, since no type\,Ia supernova has yet been detected in the radio regime in M\,82 \citep{Torres2014}, we can firmly say that SN$\,$2008iz is a core collapse supernova, in agreement with both \cite{Brunthaler2009b} and \cite{Mattila2013}. 

The position of supernova SN$\,$2008iz is estimated to be $\mathrm{\alpha_{J2000}=09^{h}55^{m}51^{s}.551\pm0^{s}.008}$, $\mathrm{\delta_{J2000}=+69^{\circ}40^{\prime}45^{\prime\prime}.792\pm}$ $\mathrm{0^{\prime\prime}.005}$
\citep{Brunthaler2009b} which is 2.5$^{\prime\prime}$ (43$\,$pc)
South-West of the photometric center of M$\,$82 based on the 2.2$\mu$m
peak \citep{Weiss2001}. The location of the supernova is heavily obscured by dust and gas, hiding the region from direct observations at optical wavelengths \citep{Weiss2001}. \cite{Brunthaler2010} estimated the extinction towards SN$\,$2008iz to be $A_{\upsilon}\sim$24.4$\,$mag, concluding that the supernova non-detections was because it exploded behind a large dusty interstellar cloud. However, using the same data and relations \cite{Mattila2013} obtained a total extinction of $\sim$48.9$\,$mag suggesting that the authors may have missed to multiply the value of N(H$_2$) by 2 to convert to hydrogen nuclei before applying the extinction relation of \cite{Guver2009}. The near-IR detection could indicate that the supernova was located in the foreground with most of the H$_2$ column density behind the site of SN$\,$2008iz \citep{Mattila2013}. 
\\
VLBI observations reveal the expansion of SN$\,$2008iz to be self-similar with an expansion velocity of $\approx$21$\,$000$\,$km$\,$s$^{-1}$ over the first 430 days
after the explosion, making it one of the fastest expanding radio
supernovae, with an expansion index, $m$, of 0.89$\pm$0.03 \citep{Brunthaler2010}.
The modeled light curve by \citep{Marchili2010} shows that synchrotron
self-absorption (SSA) is negligible, making free-free absorption (FFA)
the most significant process during the supernova expansion.
\\
Radio monitoring of SN$\,$2008iz with VLA and VLBI has been ongoing
on a regular basis since its discovery. We describe the monitoring
and data reduction process in section 2. In section 3 we present multi-frequency
radio observations from 36 to 2167$\,$days after the explosion. In
section 4 we discuss the variations of the spectral index $\alpha$,
equipartition minimum energy and magnetic fields, the evolution of
the magnetic field evolution B$_{\mathrm{eq}}$ and its amplification.
We present our summary in section 5.

\section{Observations and data calibration}
The radio monitoring campaign conducted with the VLA, at 1.4, 4.8,
8.4, 22, and 43$\,$GHz to trace the evolution of the radio emission
from SN$\,$2008iz has been ongoing since its discovery in 2008. The
first series of observations is archival and corresponds to the data
published by \citep{Brunthaler2009b,Brunthaler2010}. It consists
of six observations (epochs 1$\,$-$\,$6 in table \ref{table1})
obtained between 2008 March 24 (day 36) and 2009 September 19 (day
580). For details on the observations of those epochs, the reader
is referred to their publications. For the later observations, the
data were taken in the standard continuum observation mode with a total
bandwidth of 128$\,$MHz, each in dual circular polarisation. During
these epochs, observations were done in VLA configurations A, AB,
B, and C. Flux density measurements were derived using calibrator 3C\,48.
The flux calibrator was observed for a total time of 2 minutes in
each observation. The observation used a switching cycle of six minutes,
spending on average 1 minute on the phase calibrator J1048+7143 and
5 minutes on M$\,$82. The cycles were repeated 5 times over the observations,
yielding an integration time of $\sim$25 minutes on M$\,$82 at each frequency. 

The last epoch of VLA observations on 2014 January 23 was obtained
during a confirmation search of a newly discovered Type Ia supernova,
SN$\,$2014j at C and K bands under the observation code TOBS0008
at configuration AB. Data were taken in the standard continuum
observation mode with a total bandwidth of 128$\,$MHz, each in dual
circular polarisation. The flux calibrator 3C\,48 was observed for a
total time of 3 minutes. The observation used a switching cycle of
9 minutes, spending an average time of 30$\,$seconds on the phase
calibrator J1048+7143 and 8 minutes on M$\,$82. The cycle was repeated
4 times over the observation, yielding an integration time of $\sim$32
minutes on M$\,$82 at both frequencies.

The VLA observations in this work are matched by high resolution VLBI
observations at 1.6, 4.8, and 8.4$\,$GHz. These were taken with a
network consisting of the Very Long Baseline Array (VLBA), the Green
Bank 100 meter telescope and the Effelsberg 100 meter telescope leading
to a maximum resolution of 4.3 milliarcseconds (mas) at 1.6$\,$GHz,
1.2$\,$mas at 4.8$\,$GHz, and 0.8$\,$mas at 8.4$\,$GHz. SN$\,$2008iz
was observed 15 times between 2009 October and 2013 January in dual
polarisation. The total bandwidth of most datasets is 64$\,$MHz,
although two datasets have a bandwidth of 32$\,$MHz and three have
a bandwidth 260$\,$MHz wide. The fluxes were calibrated using system
temperature measurements and standard gain curves, and the AGN in
M$\,$81 (M$\,$81{*}) was used as a phase reference calibrator. The calibrator
J1048+7143 was observed several times during each observation run,
each time for 90 to 120 seconds. SN$\,$2008iz and M$\,$81{*} were observed
in turns for 45 to 55 seconds at 1.6$\,$GHz and $\sim$80 seconds
at 4.8, and 8.4$\,$GHz during each observation run. 

The VLA and VLBI data were reduced using standard packages within
the AIPS (Astronomical Image Processing Software) of the National
Radio Astronomy Observatory (NRAO). To ascertain the late-time enhancement
of flux values, which are much higher than expected from the supernova
radio light curve, the VLA data was run using ``CASA-EVLA\_pipeline1.1.3''
of NRAO for all epochs after 2009 September 19. The results from both
data reduction procedures are in agreement.

\section{Results}
Tables \ref{table1} and \ref{table1b} contain the complete VLA
and VLBI logs of radio flux measurements of SN$\,$2008iz. The first
column in each table lists the observation dates. The second column
lists the number of days elapsed since explosion, which is estimated
to be 2008 February 18 ($\pm$6) \citep{Marchili2010} while the rest
of the columns lists the integrated/peak flux density or source size
at different frequencies. To minimize the contribution of the diffuse
emission in M$\,$82 on the supernova measurements, we use only baselines
longer than 30$\,$k$\mathrm{\lambda}$ in the imaging. The AIPS task
`JMFIT' is used to determine the flux by fitting a 2D Gaussian to
the identified compact supernova source. The flux errors of the VLA
observations are derived by adding in quadrature the formal errors
from the 2D Gaussian, a 5\% systematic error, and a small random error
derived as the difference between peak and integrated flux densities.
The errors of the VLBI 1.6$\,$GHz fluxes are derived in a similar
way; adding in quadrature the formal errors from 2D Gaussian fit to
the sources, and a 9\% systematic error. However, the supernova is
resolved at 4.8 and 8.4$\,$GHZ in the VLBI data, thus their integrated
fluxes cannot be derived in the same way as for the other observations.
In table \ref{table1b}, we list the peak intensities at 4.8 and 8.4$\,$GHz.

\newpage

\subsection{Light curve}\label{lightcurve}
\begin{figure}[h!]
  \resizebox{0.9\hsize}{!}{\includegraphics{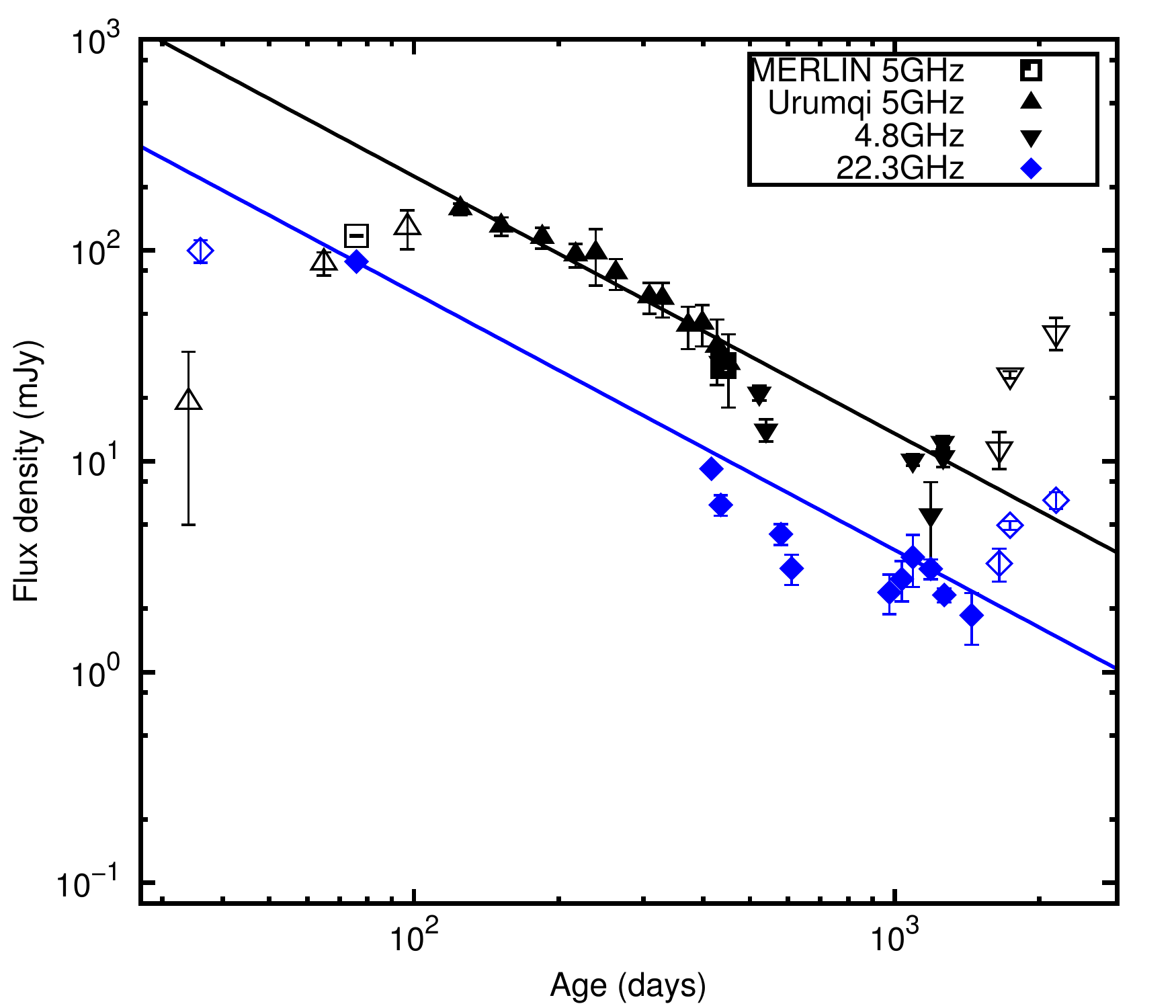}}
  \caption{Fits to the light curve of the optically-thin regime of SN$\,$2008iz
at 5 and 22$\,$GHz. $\beta$ and K$_{0}$ (see equation 1) are derived
from the fit to the data over the range of 100-1500 days for 5$\,$GHz
and 60 -- 1500 days for 22$\,$GHz. The fitted data are shown as filled
symbols, while the data not used to fit are shown as open symbols.
The MERLIN 5$\,$GHz data were taken from \citet{Beswick2009}, and
the Urumqi 5$\,$GHz from \citet{Marchili2010}. The rest of the
data are tabulated in Table \ref{table1}.}
  \label{opticalthin5_22GHz}
\end{figure}

\begin{figure}[H]
  \resizebox{0.9\hsize}{!}{\includegraphics{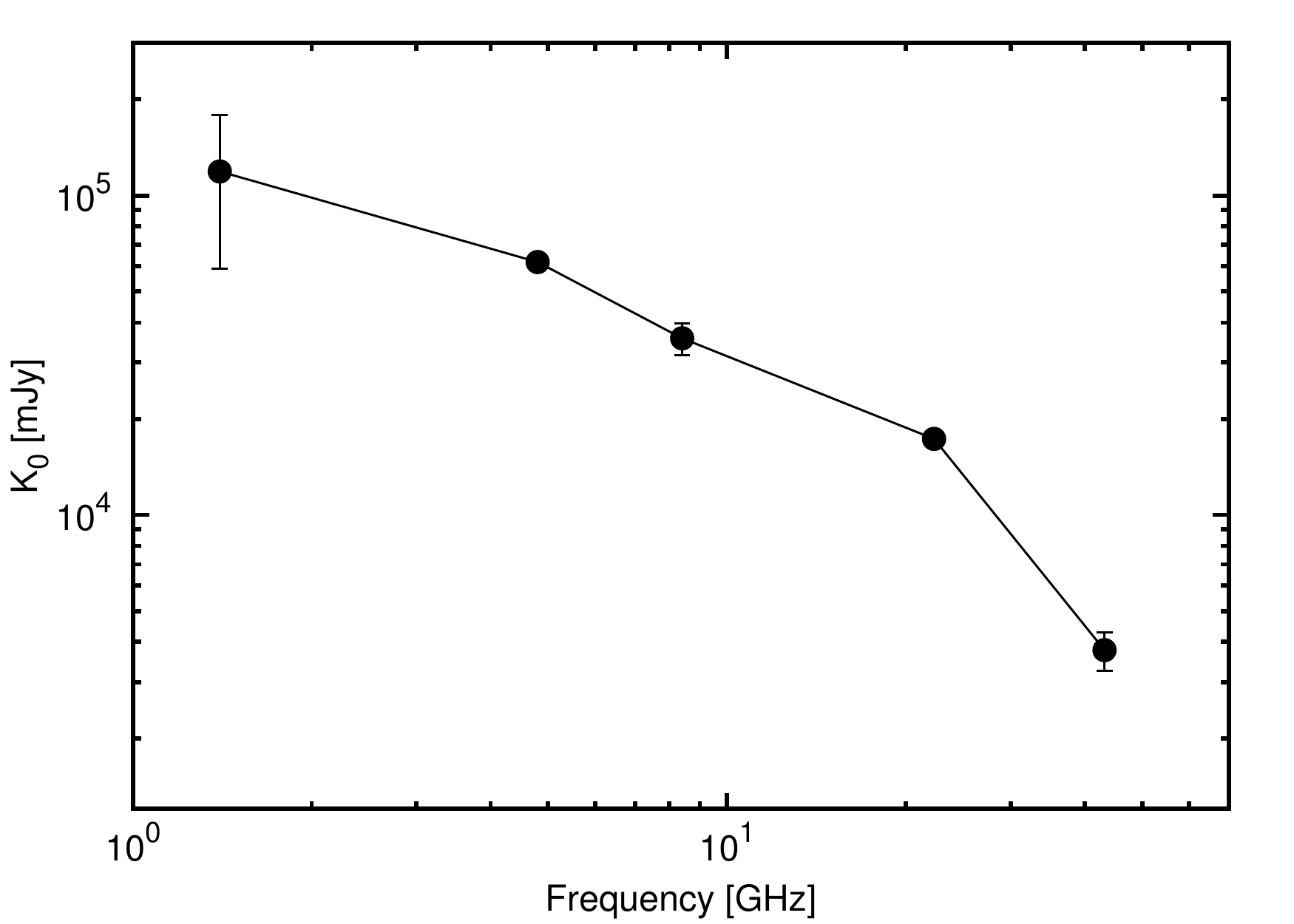}}
  \caption{The normalized average spectrum of the variation of K$_O$  with frequency for the optically-thin power-law model.}
  \label{KovsNu}
\end{figure}

\begin{figure}[h!]
  \resizebox{0.9\hsize}{!}{\includegraphics{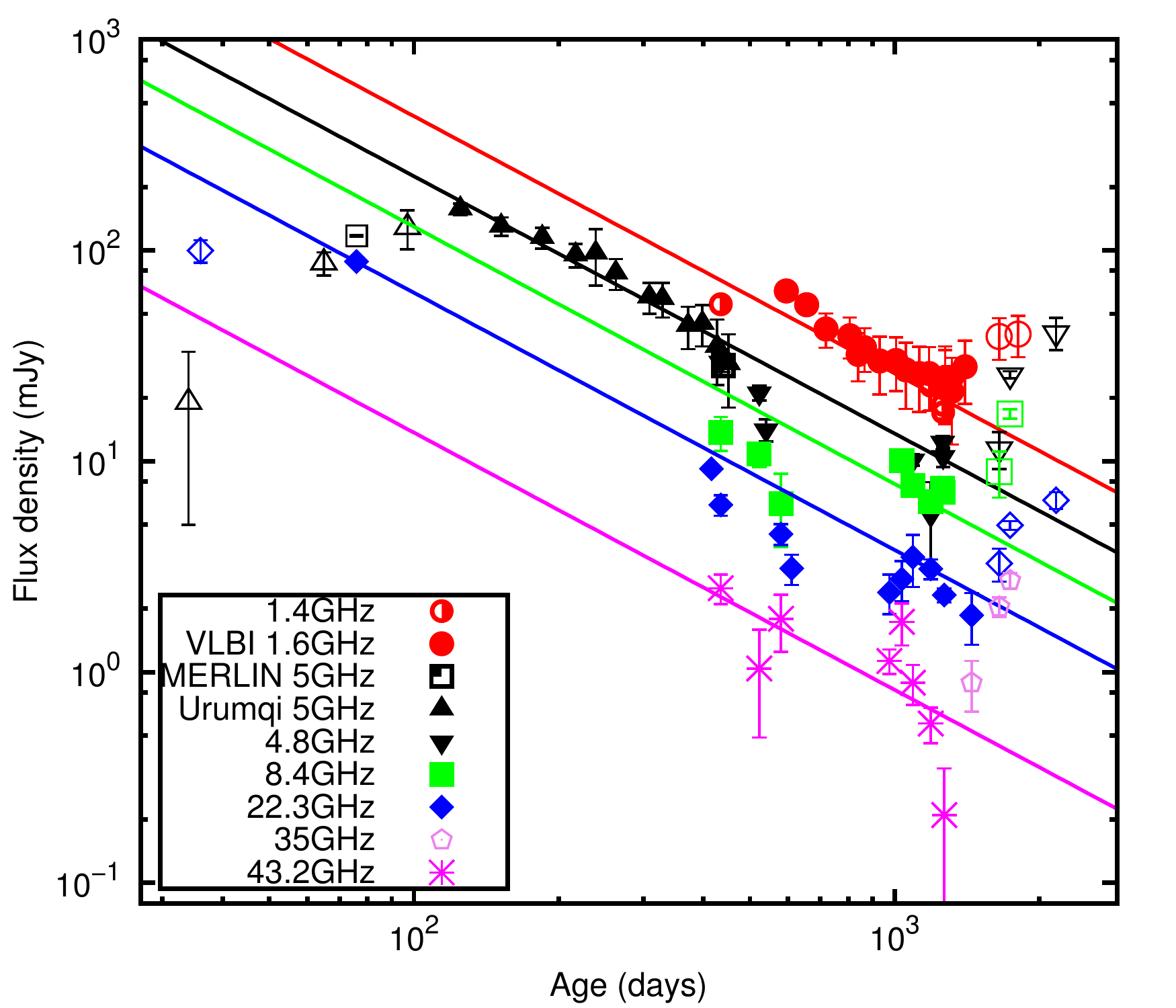}}
  \caption{A combined multi-frequency light curve for the optically-thin regime
of SN$\,$2008iz. The fitted data are shown as filled symbols, while
the data not used to fit are shown as open symbols. The data sources
are indicated in Fig \ref{opticalthin5_22GHz}, while the 1.6$\,$GHz
VLBI data are tabulated in Table \ref{table1b}.The lines represent
simple power-law fits to the data at the different frequencies as
color coded in the legend.}
  \label{opticalthin}
\end{figure}

The light curve of SN$\,$2008iz is well fit by a power-law of the
form, 

\begin{equation}
S=K_{0}\left(\frac{t-t_{0}}{1\, \mathrm{day}}\right)^{\beta}\label{eq:sn2008izeqn1}
\end{equation}

\noindent especially in the well-sampled optically thin regime, i.e., 5 and 22$\,$GHz (see Fig. \ref{opticalthin5_22GHz}). $K_{0}$ is a flux scaling
factor which does not change the overall behaviour of the fit, (t$-$t$_{0}$)
is the time elapsed after the explosion date, while $\beta$ is the
flux decay index. The obtained fit results (see Fig. \ref{opticalthin5_22GHz})
with $\beta$ and $K_{0}$ as free parameter at 5$\,$GHz (100 --
1500 days) are $\mathrm{K}_{0\,5\, \mathrm{GHz}}$= (6.29$\pm$2.22)$\times$10$^{4}$$\,$mJy
and $\beta$=$-$1.22$\pm$0.07. The time range 0 -- 100 days is excluded
because the radio emission from the supernova shock wave was still
within the optically-thick regime, while the last epochs of the data
sets have their peak intensity higher than expected from the light
curve. The obtained fit value at 22$\,$GHz (50 -- 1500 days) with
both $\beta$ and $K_{0}$ as free parameters yields $\mathrm{K}_{0\,22\,\mathrm{GHz}}$=
(1.46$\pm$0.33)$\times$10$^{4}$$\,$mJy and $\beta$=$-$1.18$\pm$0.05.
Considering the 5$\,$GHz data set, which is better sampled over the
time range than the one at 22$\,$GHz, we adopt a $\beta$ value of
$-$1.22$\pm$0.07 to be more representative. Fitting the 22$\,$GHz
data with our adopted $\beta$ value, we obtain the scaling parameter
$\mathrm{K}_{0\,22\,\mathrm{GHz}}$ to be (1.73$\pm$0.10)$\times$10$^{4}$$\,$mJy.
A fit to all frequencies (Fig. \ref{opticalthin}) using one single
$\beta$ value of -1.22$\pm$0.07 yields a different $K_{0}$ value
for each frequency. The values of $K_{0}$ are $\mathrm{K}_{0\,1.4\, GHz}$=
(1.19$\pm$0.60)$\times$10$^{5}$$\,$mJy, $\mathrm{K}_{0\,5\,\mathrm{GHz}}$=
(6.19$\pm$0.17)$\times$10$^{4}$$\,$mJy, $\mathrm{K}_{0\,8.4\,\mathrm{GHz}}$=
(3.57$\pm$0.40)$\times$10$^{4}$$\,$mJy, $\mathrm{K}_{0\,22\,\mathrm{GHz}}$=
(1.73$\pm$0.1)$\times$10$^{4}$$\,$mJy, $\mathrm{K}_{0\,43\,\mathrm{GHz}}$=
(3.77$\pm$0.52)$\times$10$^{3}$$\,$mJy. The derived scaling factor
$K_{0}$ yields a normalised average spectrum that depends on frequency
as shown in Fig.\ref{KovsNu}.

Although the simple power-law (see equation \ref{eq:sn2008izeqn1}) describes
the optically-thin regime of the light curve very well, a complete
fit to the supernova light curve model is not possible because we
have data for only very few epochs in the optically-thick regime.
For the complete radio light curve, we fit the simplified Weiler model
\citep{Weiler2007} described as

\begin{equation}
S_{(\nu)}=K_{1}\left(\frac{\nu}{5\,\mathrm{GHz}}\right)^{\alpha}\left(\frac{t-t_{0}}{1\,\mathrm{day}}\right)^{\beta}e^{-\tau},
\label{eq:sn2008izeqn2}
\end{equation}

\noindent where K$_{1}$ is a scaling factor in Jy and $\alpha$ is the spectral
index of the emission in the optically-thin regime. The exponent $\beta$
is related to the flux density drop at late epochs, normally affected
by both the spectral index, $\alpha$, and the opacity, $\tau$. The
date of explosion is denoted as t$_{0}$ and t is the epoch of observation.
The opacity of thermal electrons of the circumstellar medium (CSM)
in equation \ref{eq:sn2008izeqn2} is modeled as

\begin{equation}
\tau=K_{2}\left(\frac{\nu}{5\,\mathrm{GHz}}\right)^{-2.1}\left(\frac{t-t_{0}}{1\, \mathrm{day}}\right)^{\delta},
\label{eq:sn2008izeqn3}
\end{equation}

\noindent where K$_{2}$ is the scaling factor and $\delta$ is the absorption
decline index related to the CSM radial density profile derived as
$\delta$ =$\alpha$$-$$\beta$$-$3. The exponent $-$2.1 corresponds
to the spectral dependence of FFA by thermal ionised gas in the radio
regime.

\begin{figure}[H]
  \resizebox{0.9\hsize}{!}{\includegraphics{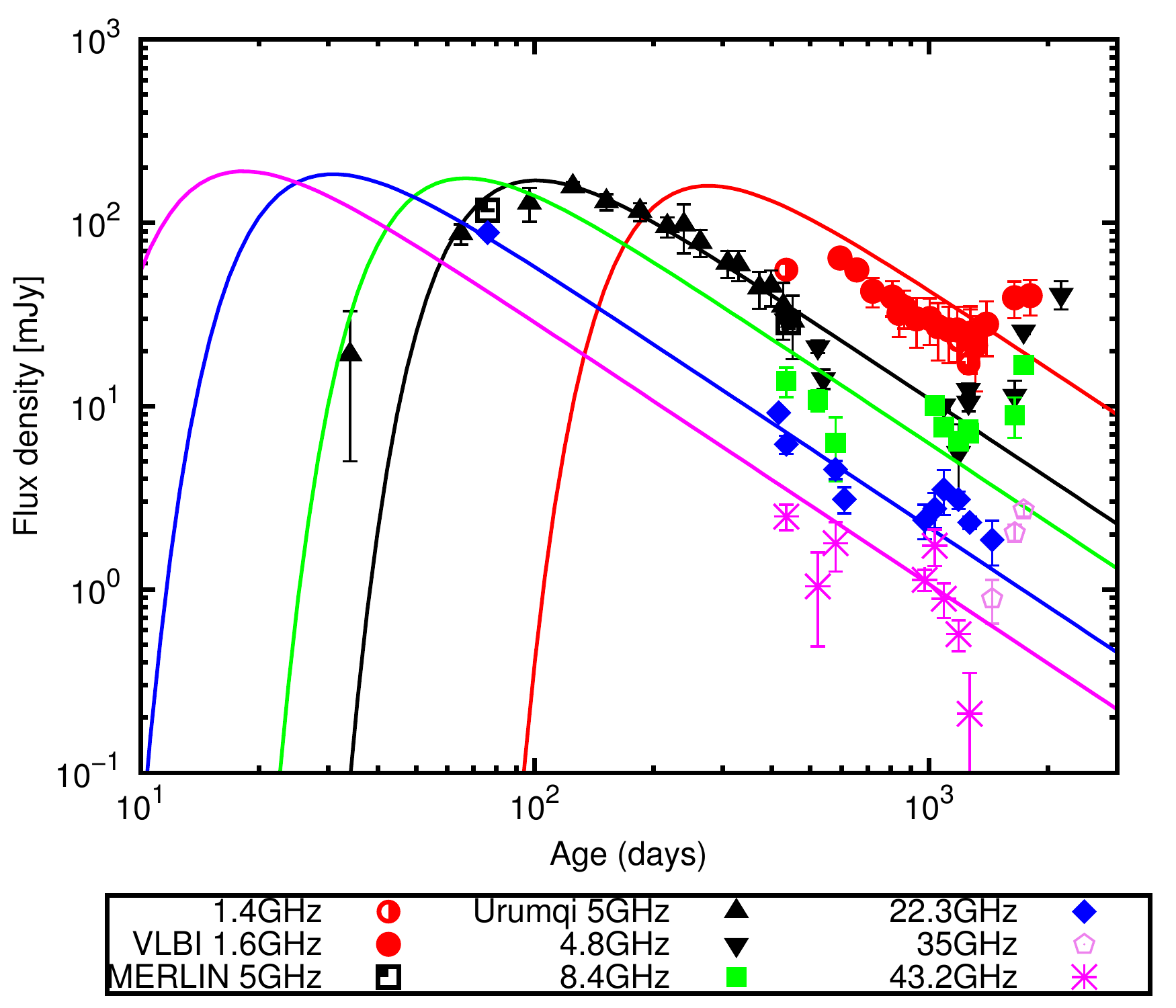}}
  \caption{The multi-frequency light curve of SN$\,$2008iz. The MERLIN 5$\,$GHz
data were taken from \citet{Beswick2009}, the Urumqi 5$\,$GHz from \citet{Marchili2010}. The rest of the data are tabulated in
table \ref{table1} and \ref{table1b}. The black line represents
a Weiler model fit to 5$\,$GHz data obtained by \citet{Marchili2010}
with the Urumqi telescope. The other lines represent light curves calculated
using the equations in the text for the other frequency data sets
as color coded in the legend.}
  \label{complete}
\end{figure}

To constrain the simplified Weiler model, we apply the best-fit parameters
of the model \ref{eq:sn2008izeqn2} and \ref{eq:sn2008izeqn3} obtained
by \citep{Marchili2010} at 5$\,$GHz (i.e. K$_{1}$=(2.14$\pm$0.04)$\times$10$^{5}$$\,$mJy,
$\beta$=$-$1.43$\pm$0.05, t$_{0}$=18 Feb 2008, $K_{2}$=(11.0$\pm$0.7)$\times$10$^{4}$$\,$mJy,
$\delta$=$-$2.65$\pm$0.10) and from \citep{Brunthaler2010} we
obtain the spectral index, $\alpha$=$-$1.08$\pm$0.08. We fit all
frequencies at the same time as shown in Fig.\ref{complete}
using $\beta$ as a free parameter, for which a value of $\beta=-$1.41$\pm$0.02
is determined. 

The model fits satisfactorily the flux density decline from day 500
except for the 1.4$\,$GHz observations that fall well below the model
fit on Fig. \ref{complete}. The lower flux densities
at 1.4$\,$GHz are also confirmed by a 3$\,$$\sigma$ LOFAR non-detection
limit of SN$\,$2008iz at a level of 0.41mJy/beam at the even lower
frequency of 154$\,$MHz \citep{Varenius2015}. The plausible explanations
could be FFA from a dense foreground screen along the line of sight.
Other effects, like low-frequency cut-off caused by Razin-Tsytovich
effect could also help explain the lower 1.4$\,$GHz flux and LOFAR
non-detection. From the same figure, the observations at 43.2$\,$GHz
drop quite fast past $\sim$1000$\,$days. This drop is most likely
related to synchrotron ageing of the emitting electrons in the shocked
CSM.

\clearpage

\subsection{Deceleration of SN\,2008iz}\label{deceleration}
\begin{figure*}
\centering
\begin{minipage}{0.25\linewidth}
  \centering
  \subcaptionbox{Day 595}
  {\includegraphics[width=\linewidth]{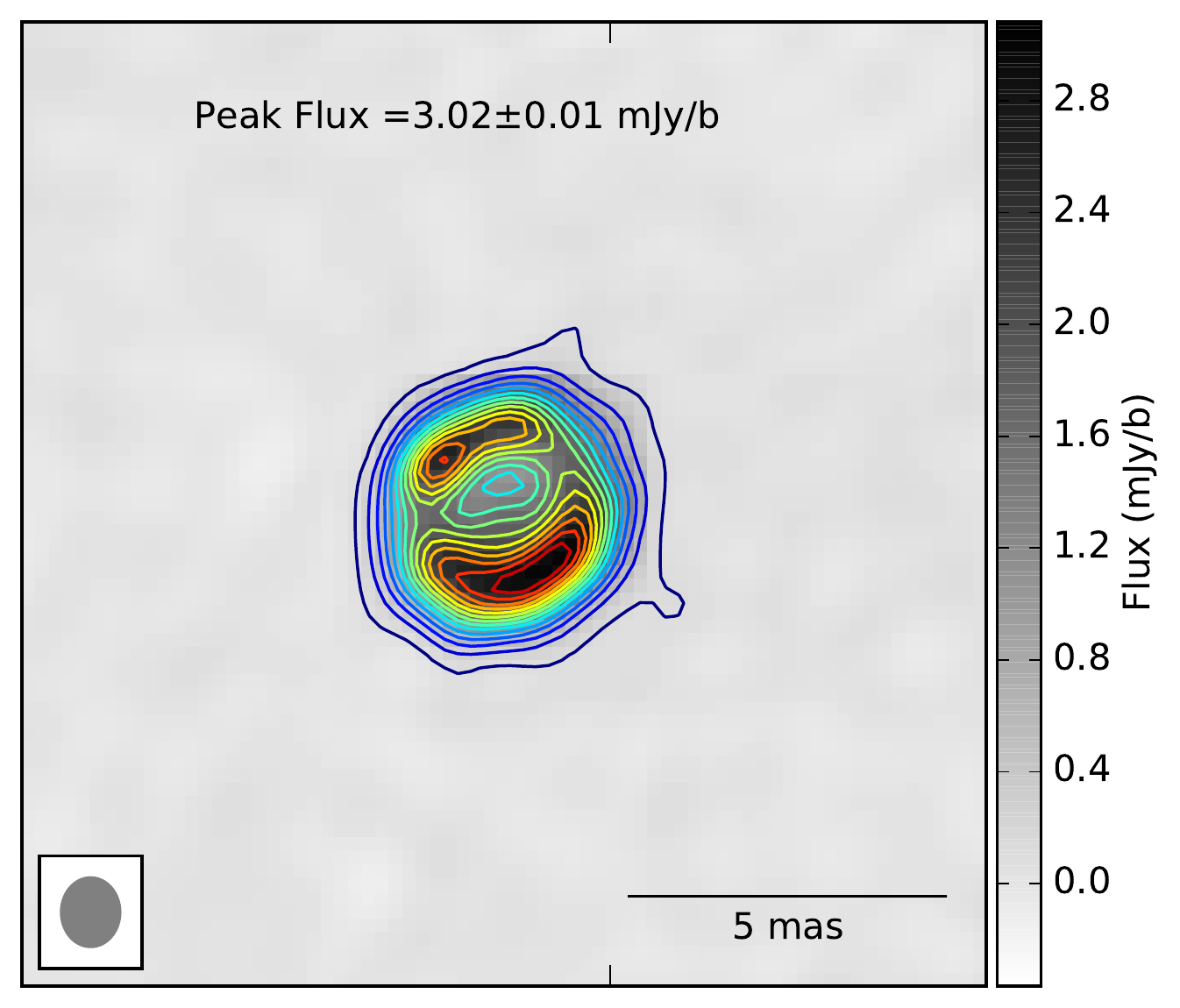}
   \label{fig:test1}}
\end{minipage}%
\begin{minipage}{0.25\linewidth}
  \centering
  \subcaptionbox{Day 656}
  {\includegraphics[width=\linewidth]{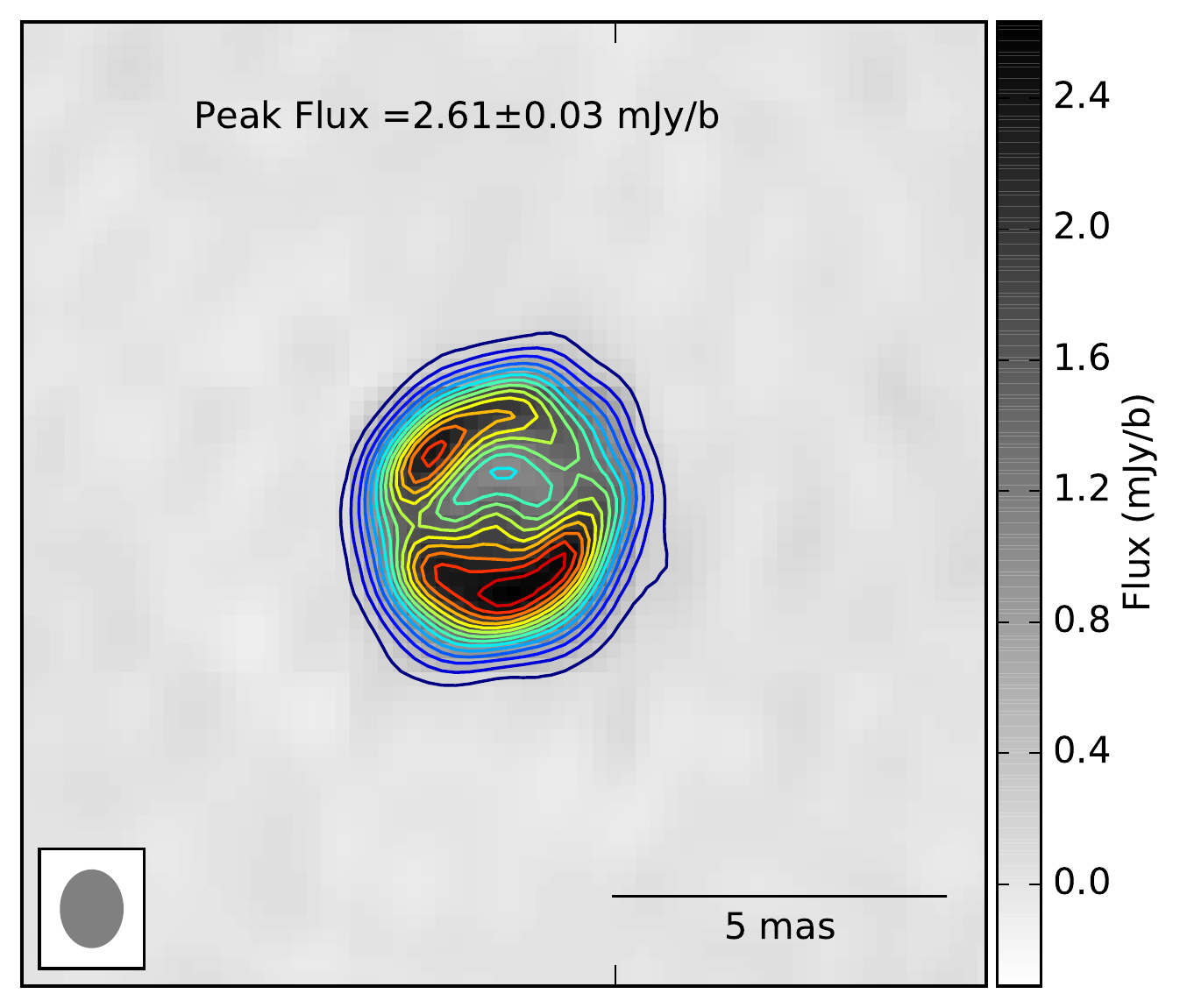}
    \label{fig:test2}}
\end{minipage}%
\begin{minipage}{0.25\linewidth}
  \centering
  \subcaptionbox{Day 720}
  {\includegraphics[width=\linewidth]{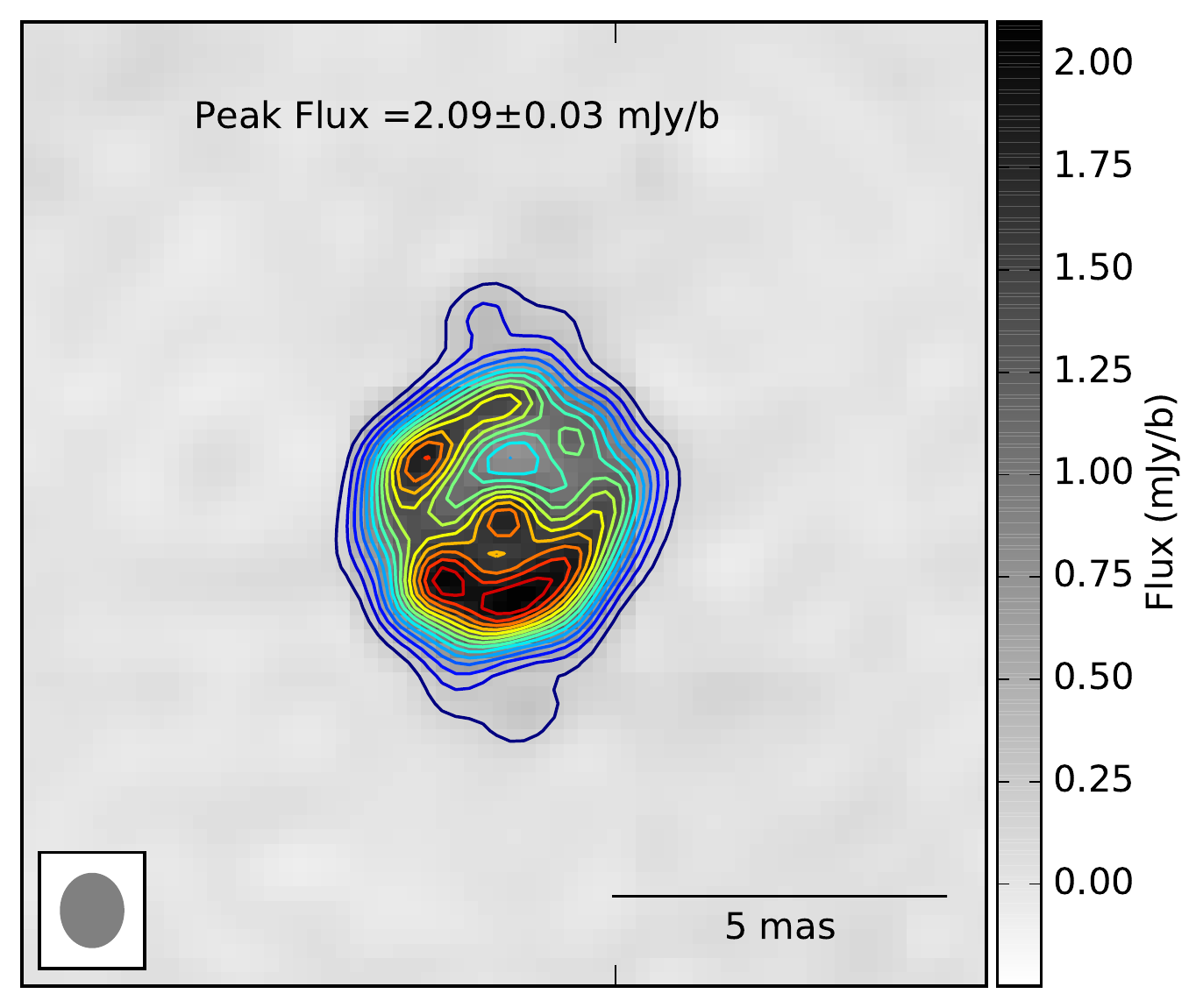}
    \label{fig:test3}}
\end{minipage}%
\begin{minipage}{0.25\linewidth}
  \centering
  \subcaptionbox{Day 807}
  {\includegraphics[width=\linewidth]{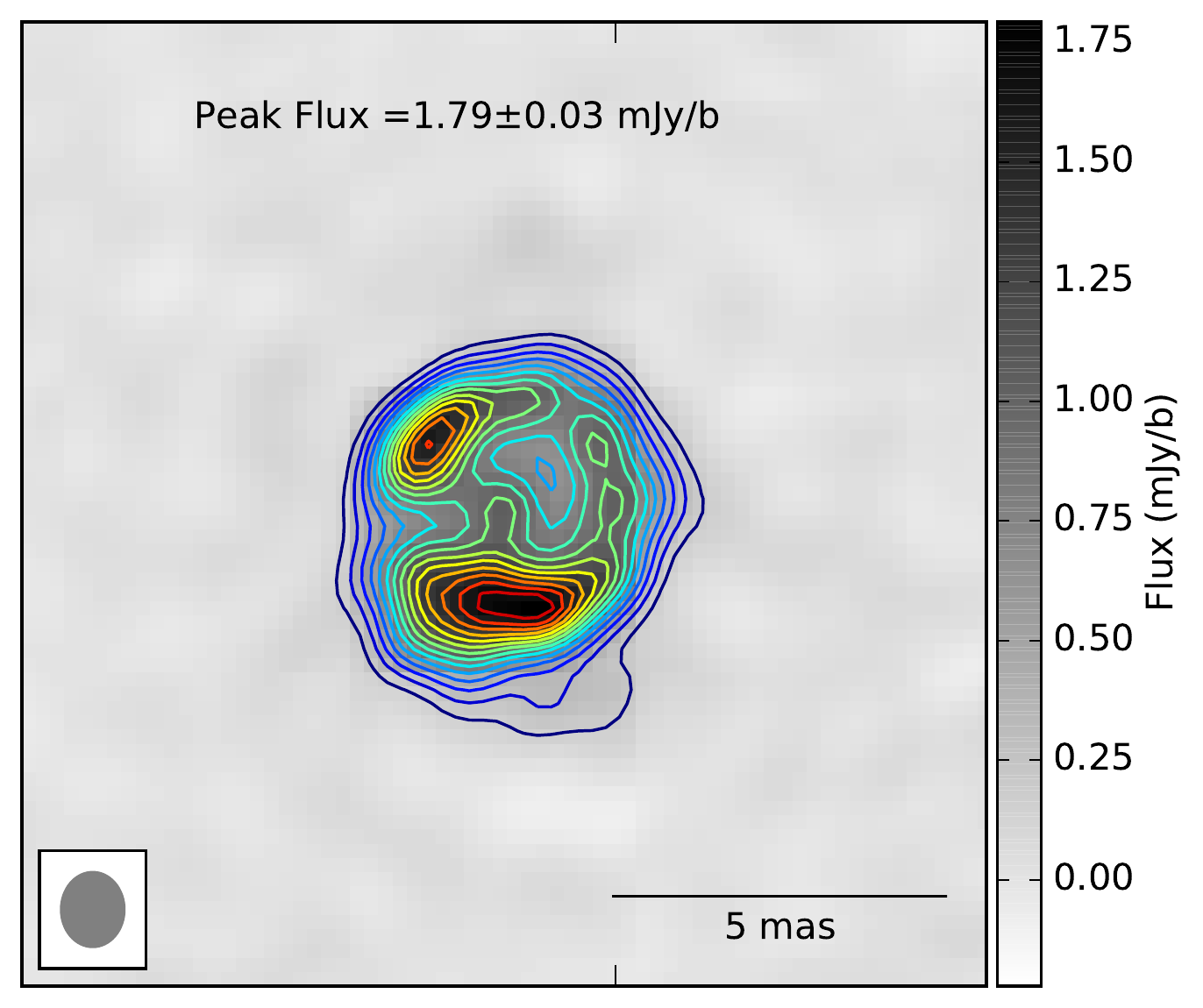}
    \label{fig:test4}}
\end{minipage}%

\begin{minipage}{0.25\linewidth}
  \centering
  \subcaptionbox{Day 866}
  {\includegraphics[width=\linewidth]{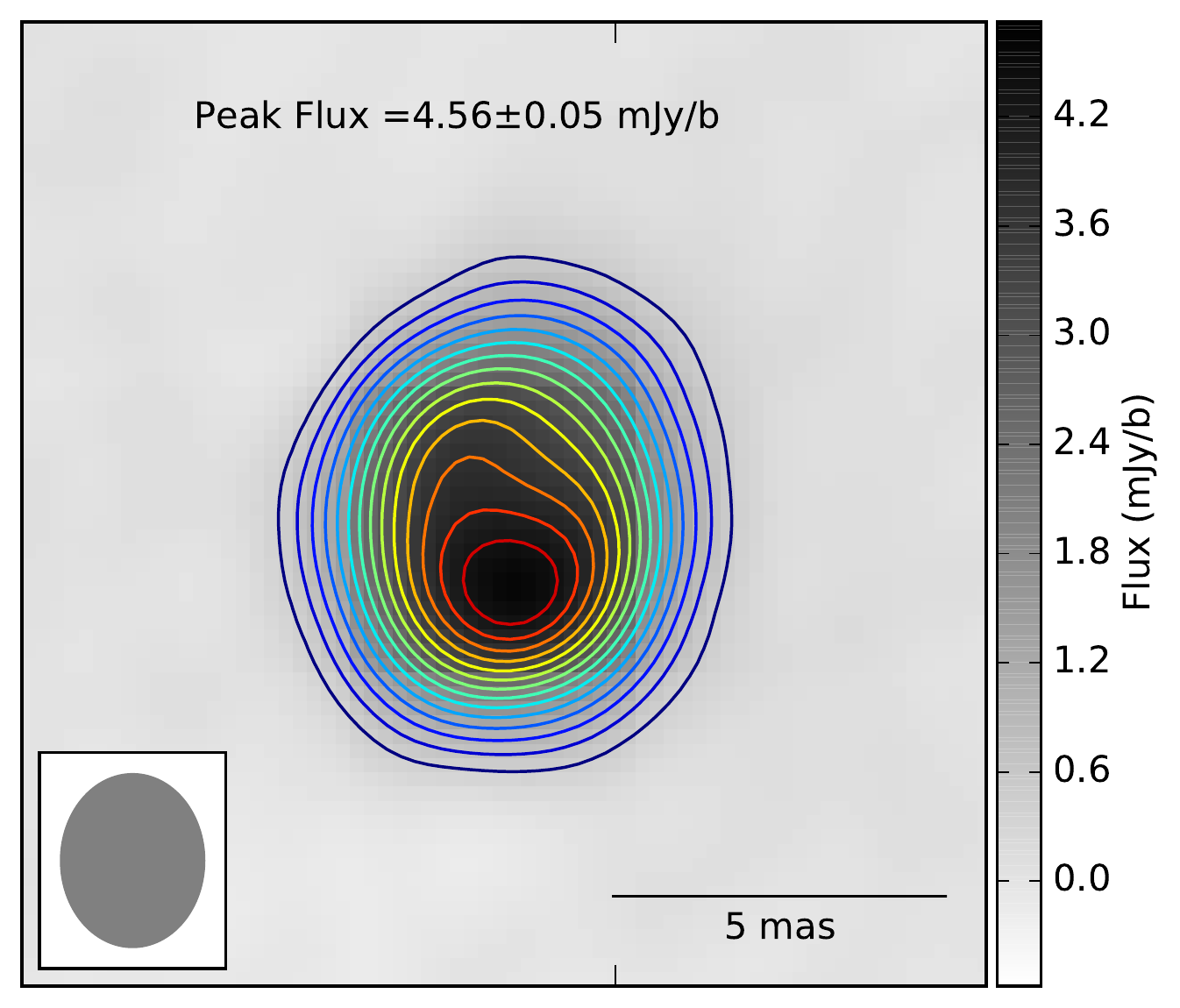}
    \label{fig:test4}}
\end{minipage}%
\begin{minipage}{0.25\linewidth}
  \centering
  \subcaptionbox{Day 931}
  {\includegraphics[width=\linewidth]{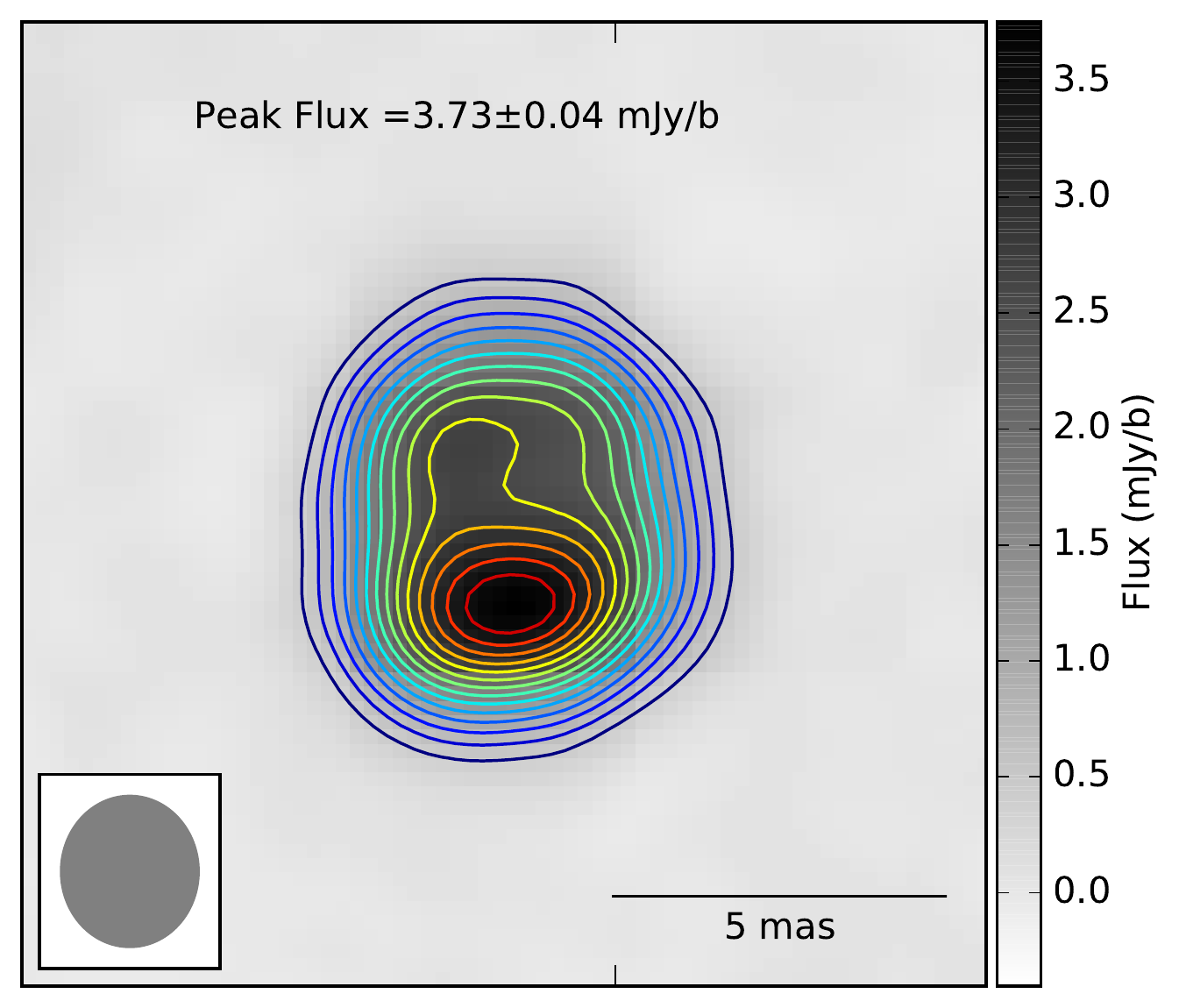}
    \label{fig:test5}}
\end{minipage}%
\begin{minipage}{0.25\linewidth}
  \centering
  \subcaptionbox{Day 1006}
  {\includegraphics[width=\linewidth]{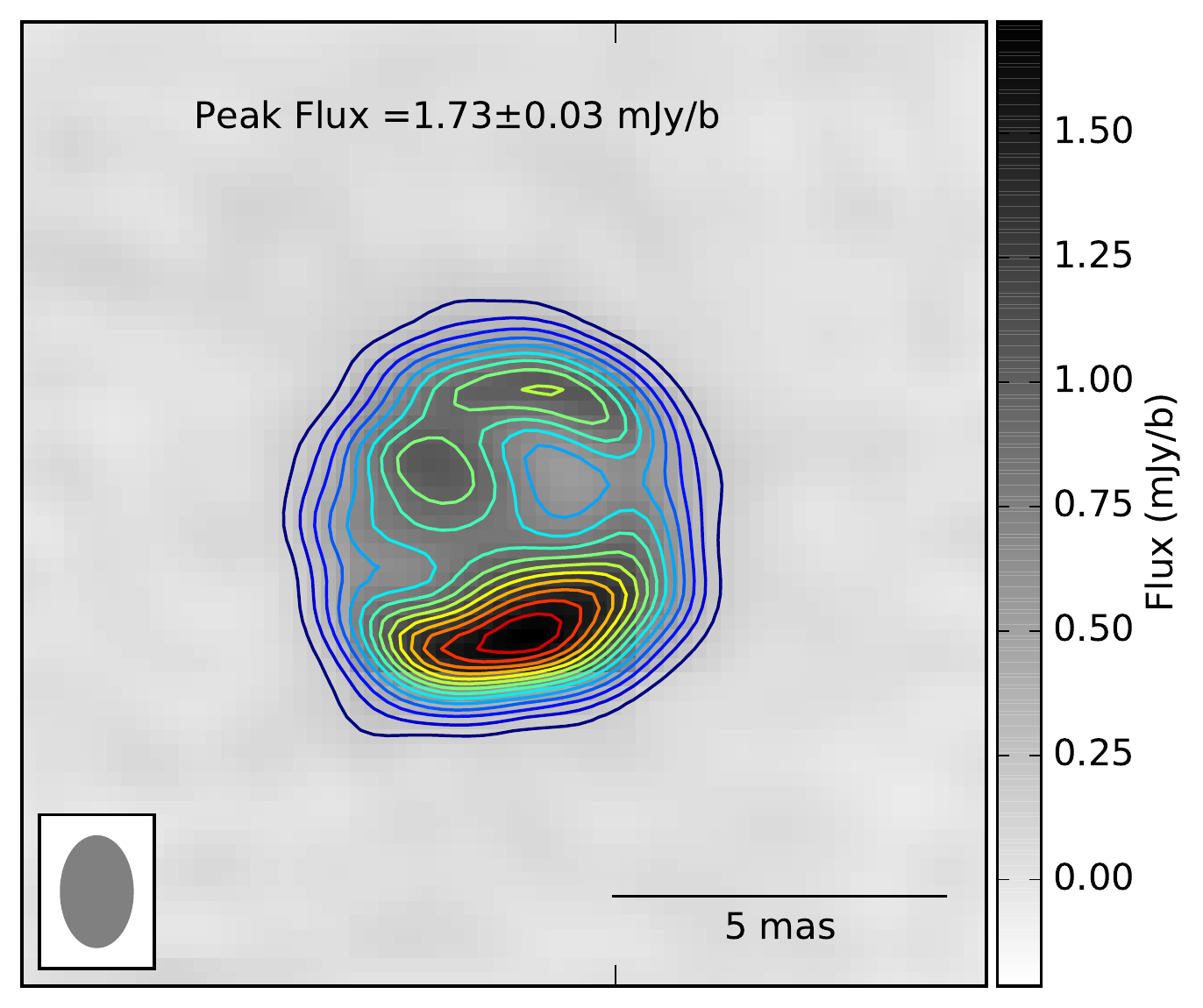}
    \label{fig:test6}}
\end{minipage}%
\begin{minipage}{0.25\linewidth}
  \centering
  \subcaptionbox{Day 1055}
  {\includegraphics[width=\linewidth]{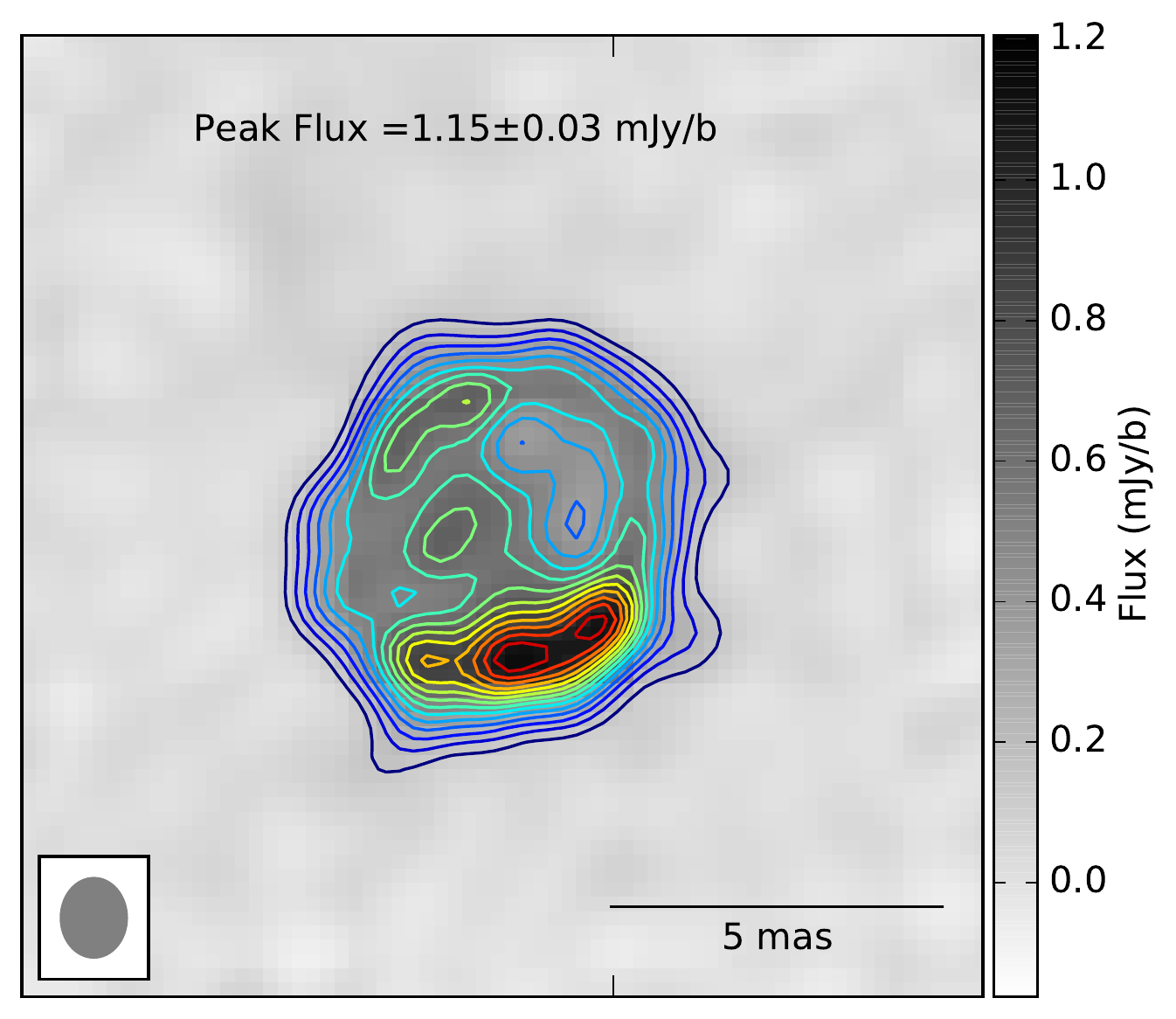}
    \label{fig:test7}}
\end{minipage}%

\begin{minipage}{0.25\linewidth}
  \centering
  \subcaptionbox{Day 1125}
  {\includegraphics[width=\linewidth]{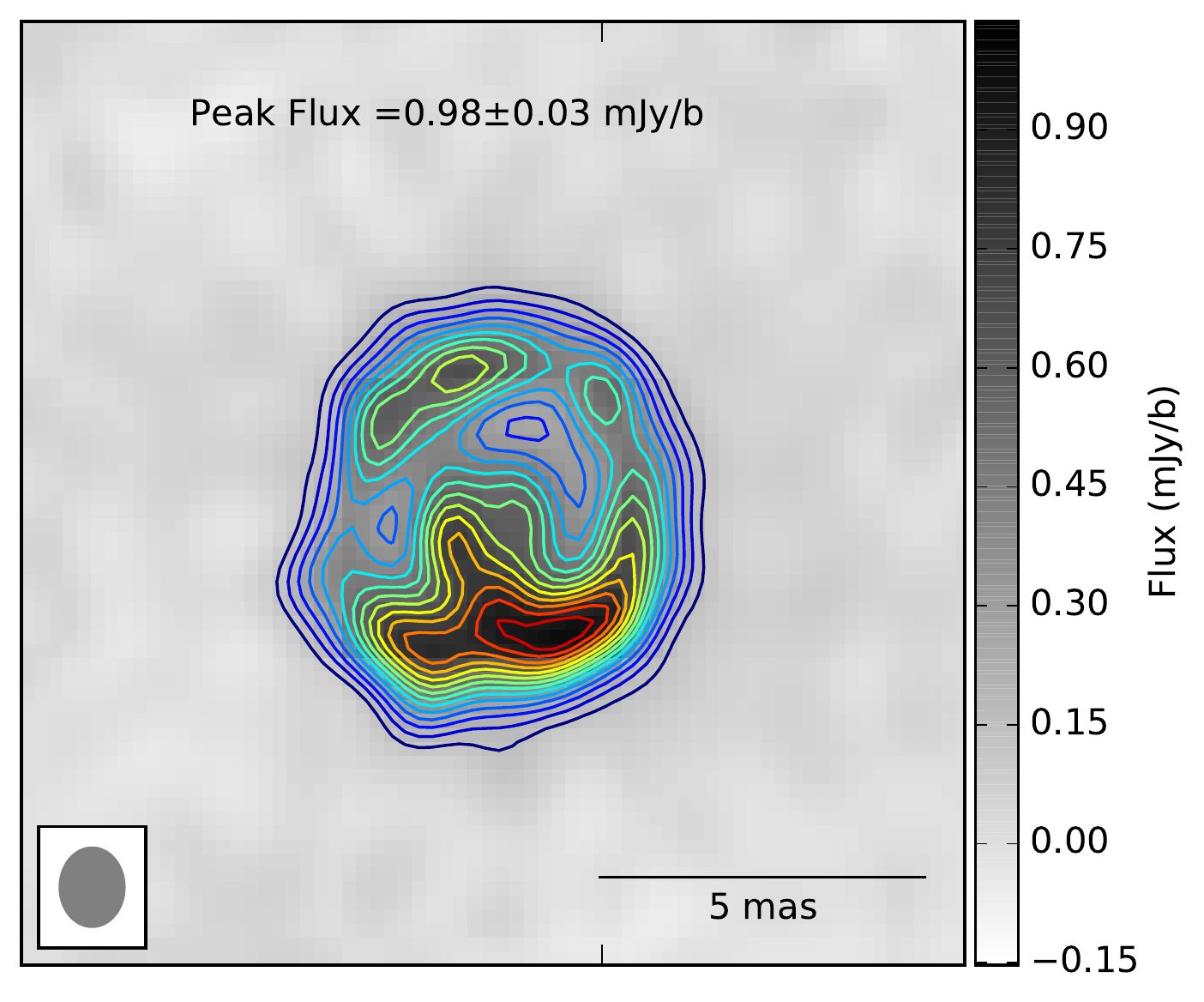}
    \label{fig:test8}}
\end{minipage}%
\begin{minipage}{0.25\linewidth}
  \centering
  \subcaptionbox{Day 1180}
  {\includegraphics[width=\linewidth]{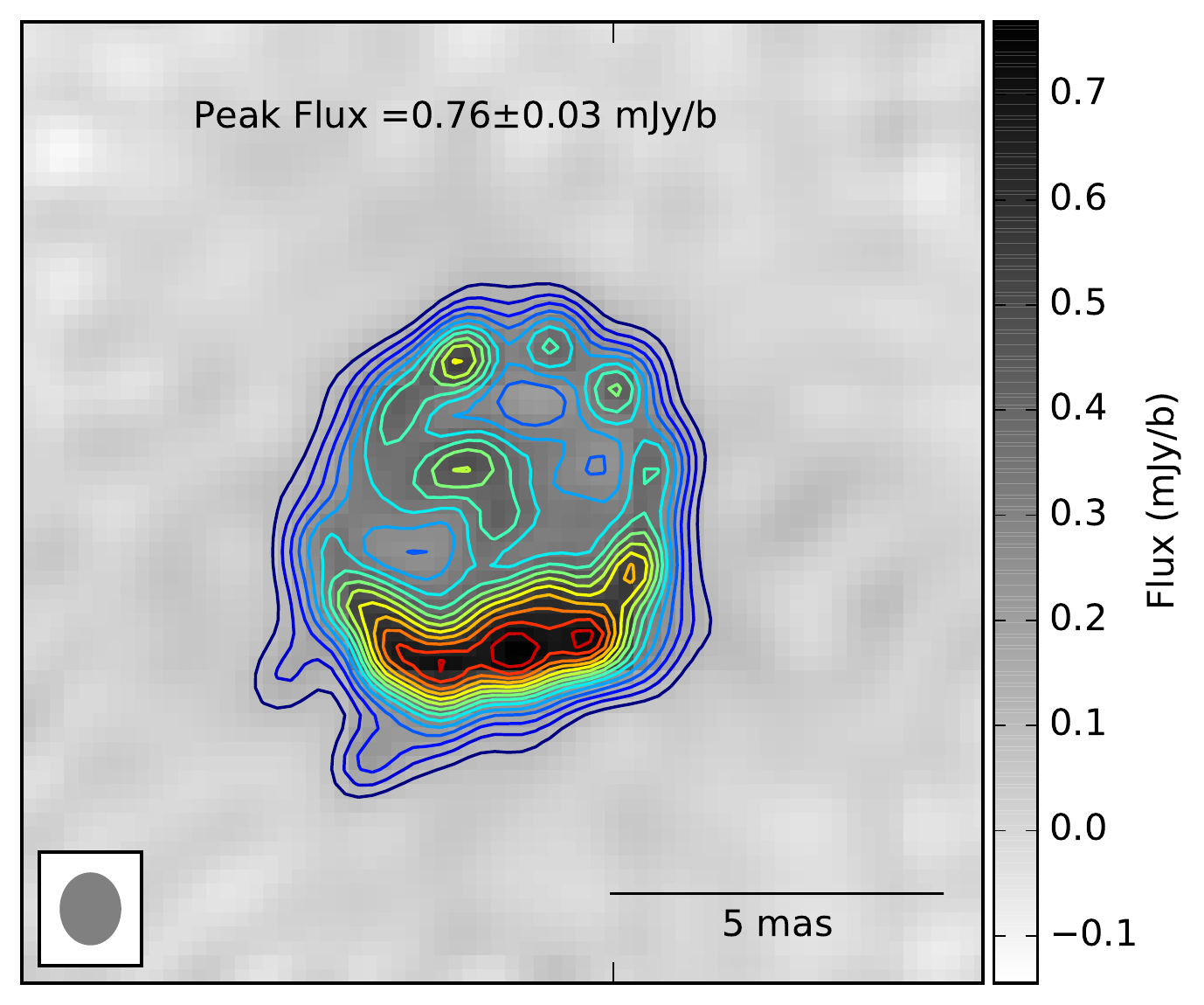}
    \label{fig:test9}}
\end{minipage}%
\begin{minipage}{0.25\linewidth}
  \centering
  \subcaptionbox{Day 1314}
  {\includegraphics[width=\linewidth]{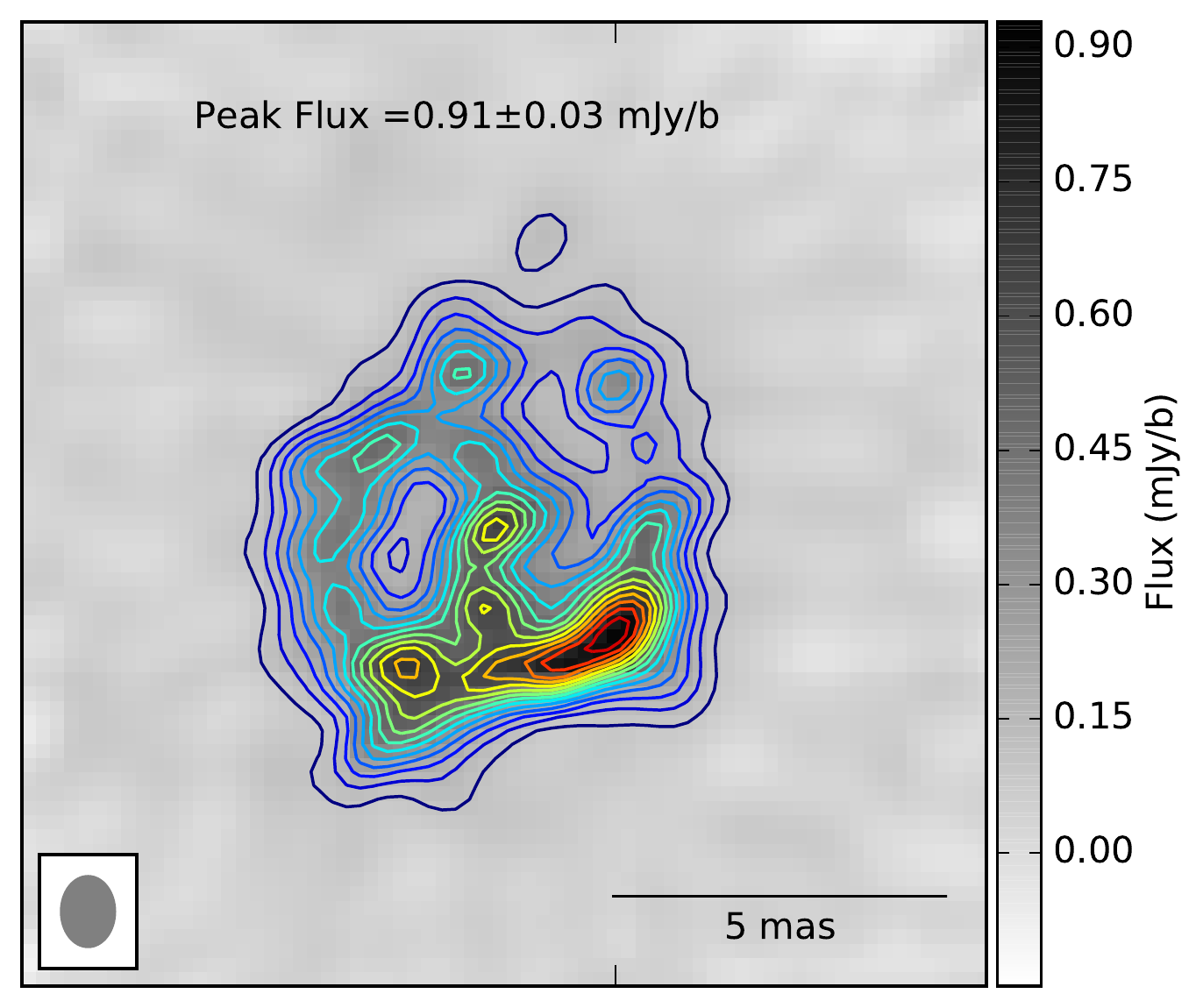}
    \label{fig:test10}}
\end{minipage}%
\begin{minipage}{0.25\linewidth}
  \centering
  \subcaptionbox{Day 1401}
  {\includegraphics[width=\linewidth]{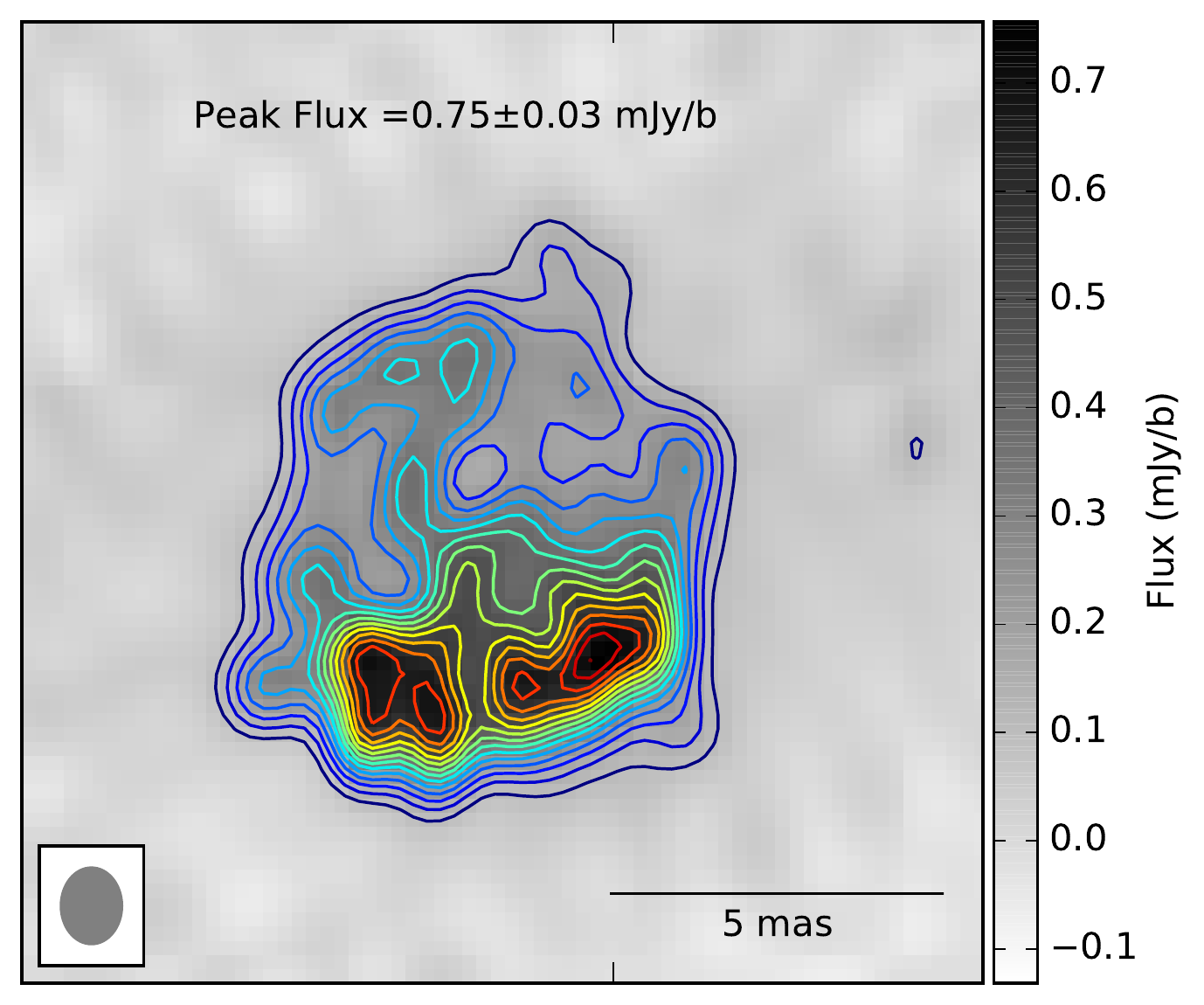}
    \label{fig:test11}}
\end{minipage}%

\begin{flushleft}
\begin{minipage}{0.25\linewidth}
  \centering
  \subcaptionbox{Day 1650}
  {\includegraphics[width=\linewidth]{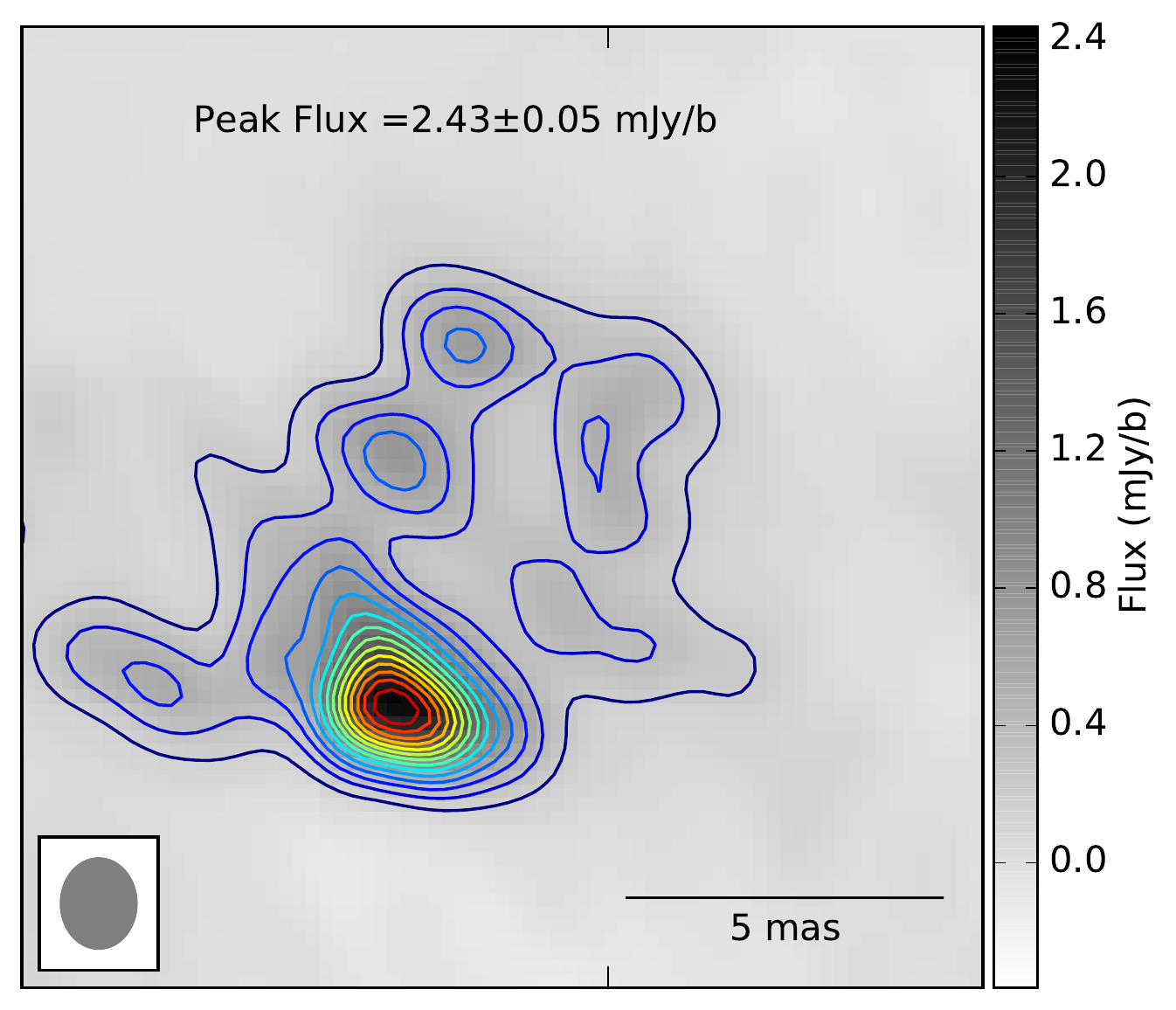}
    \label{fig:test12}}
\end{minipage}%
\begin{minipage}{0.25\linewidth}
  \centering
  \subcaptionbox{Day 1806}
  {\includegraphics[width=\linewidth]{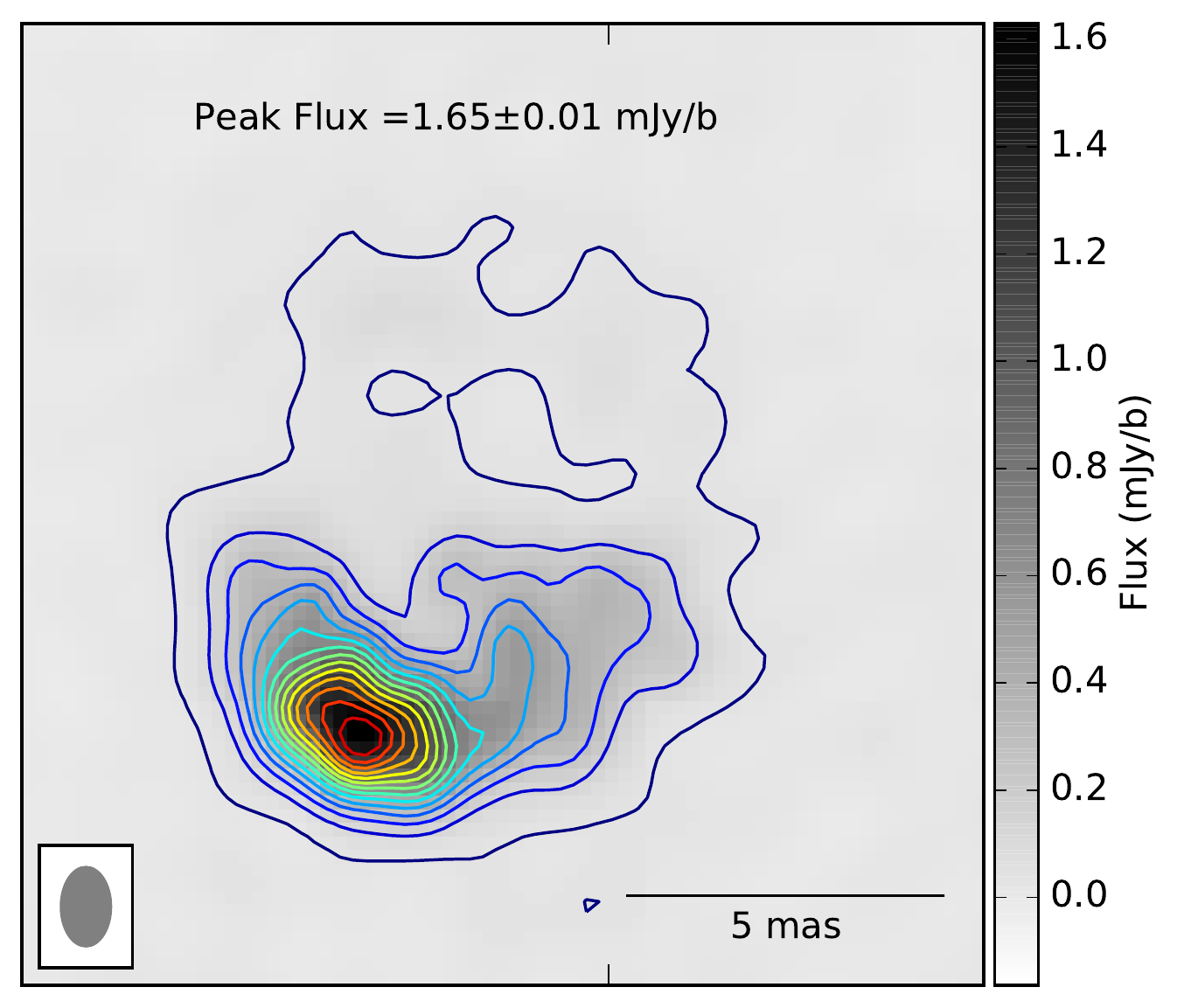}
    \label{fig:test13}}
\end{minipage}%
\end{flushleft}

\caption{The natural weighted VLBI images of the expanding shell produced by
supernova SN$\,$2008iz at 4.8$\,$GHz. They cover the observation
periods between 2009 October and 2013 January as summarized in Table$\,$\ref{table1b}. The figures show the 4.8$\,$GHz linear gray scale images overlaid
with their contour maps that scale to the peak intensity at each epoch
(listed in each panel). All image contours correspond to 6\% of the
peak intensity (lowest contour) and subsequent increases by 6\%. The
images have been convolved with the natural weighting beam, shown
at the bottom left of each panel.}
  \label{vlbi_cband}
\end{figure*}

\begin{figure*}
\centering
\begin{minipage}{0.25\linewidth}
  \centering
  \subcaptionbox{Day 595}
  {\includegraphics[width=\linewidth]{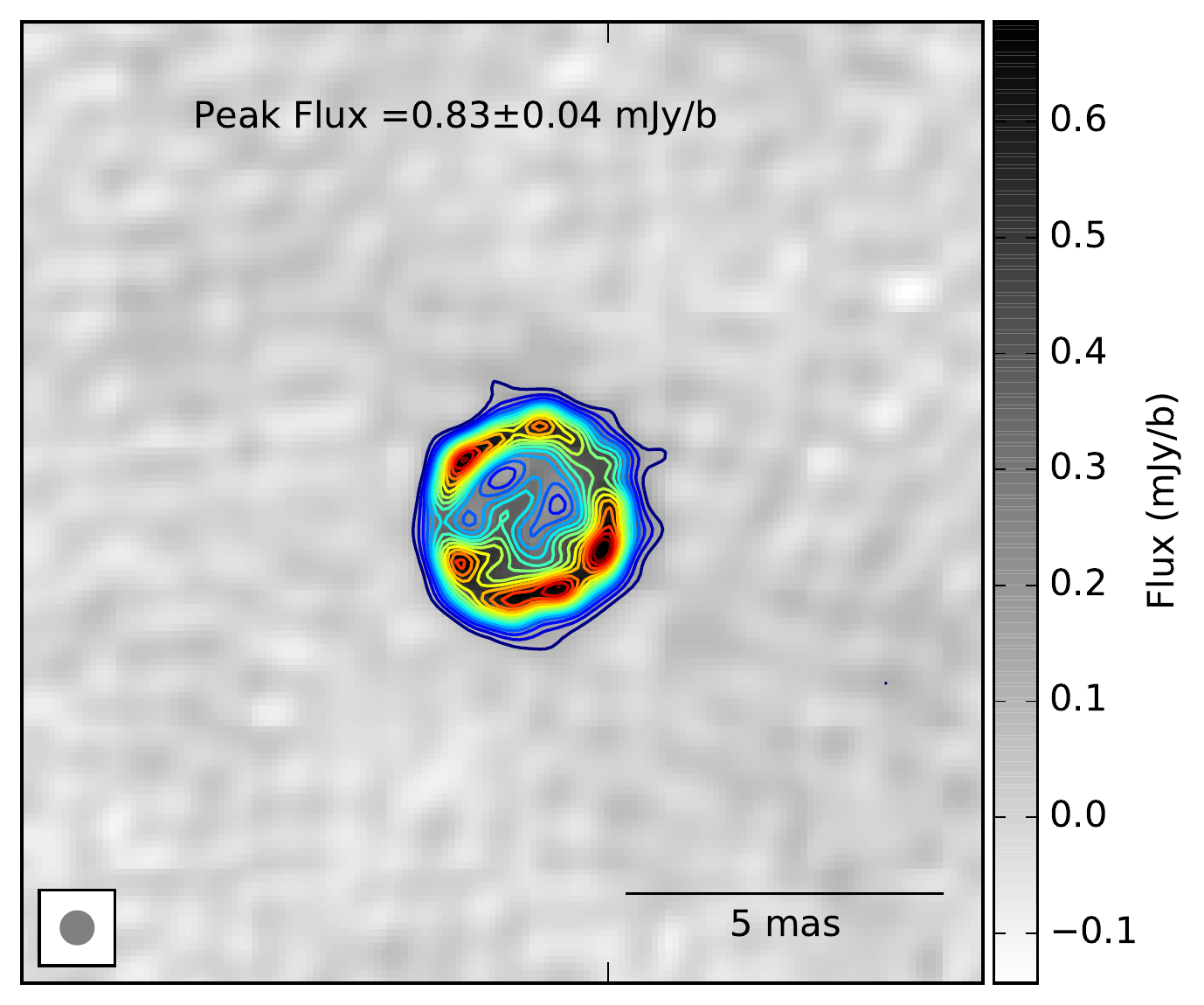}
   \label{fig:test01}}
\end{minipage}%
\begin{minipage}{0.25\linewidth}
  \centering
  \subcaptionbox{Day 656}
  {\includegraphics[width=\linewidth]{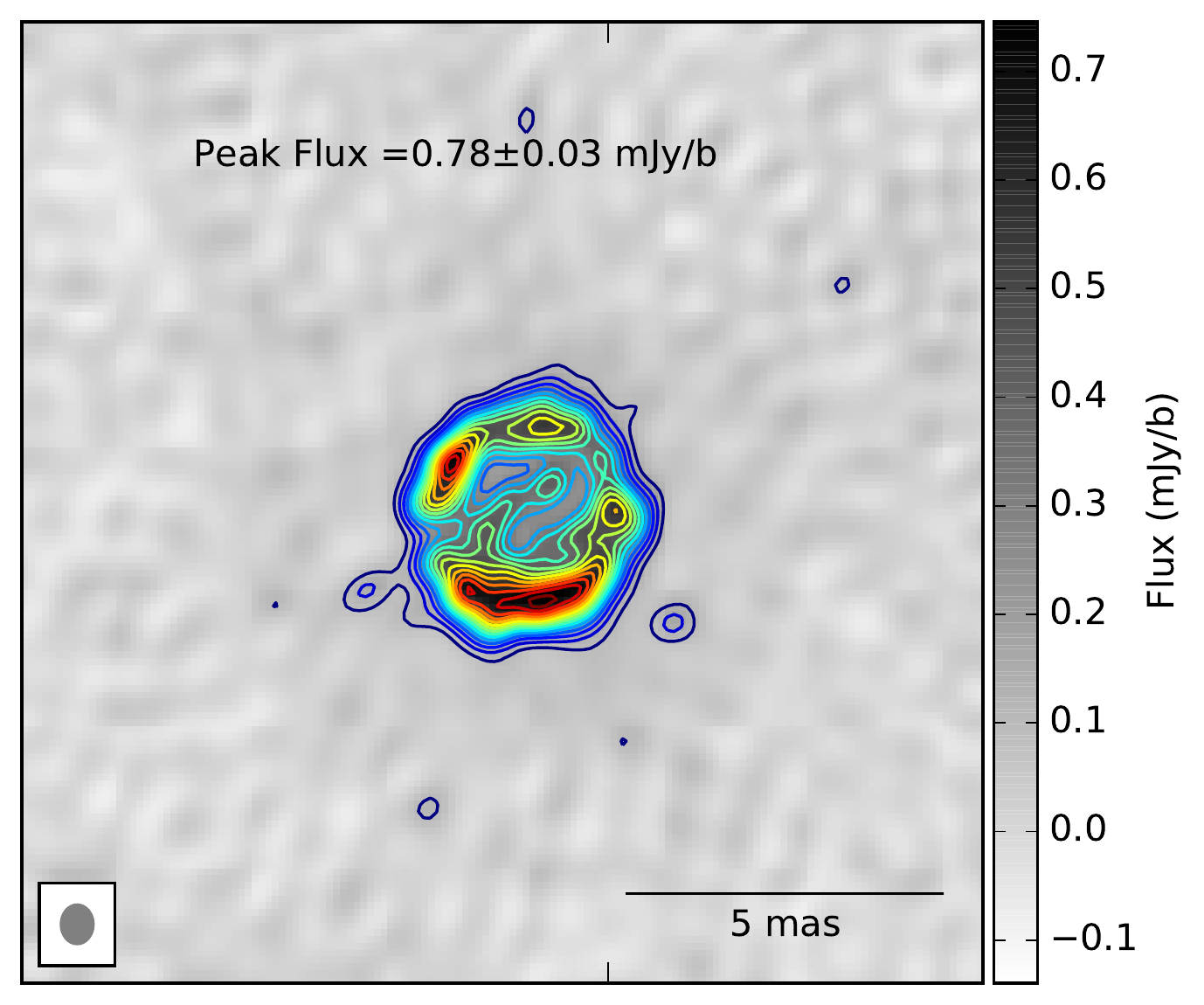}
    \label{fig:test02}}
\end{minipage}%
\begin{minipage}{0.25\linewidth}
  \centering
  \subcaptionbox{Day 720}
  {\includegraphics[width=\linewidth]{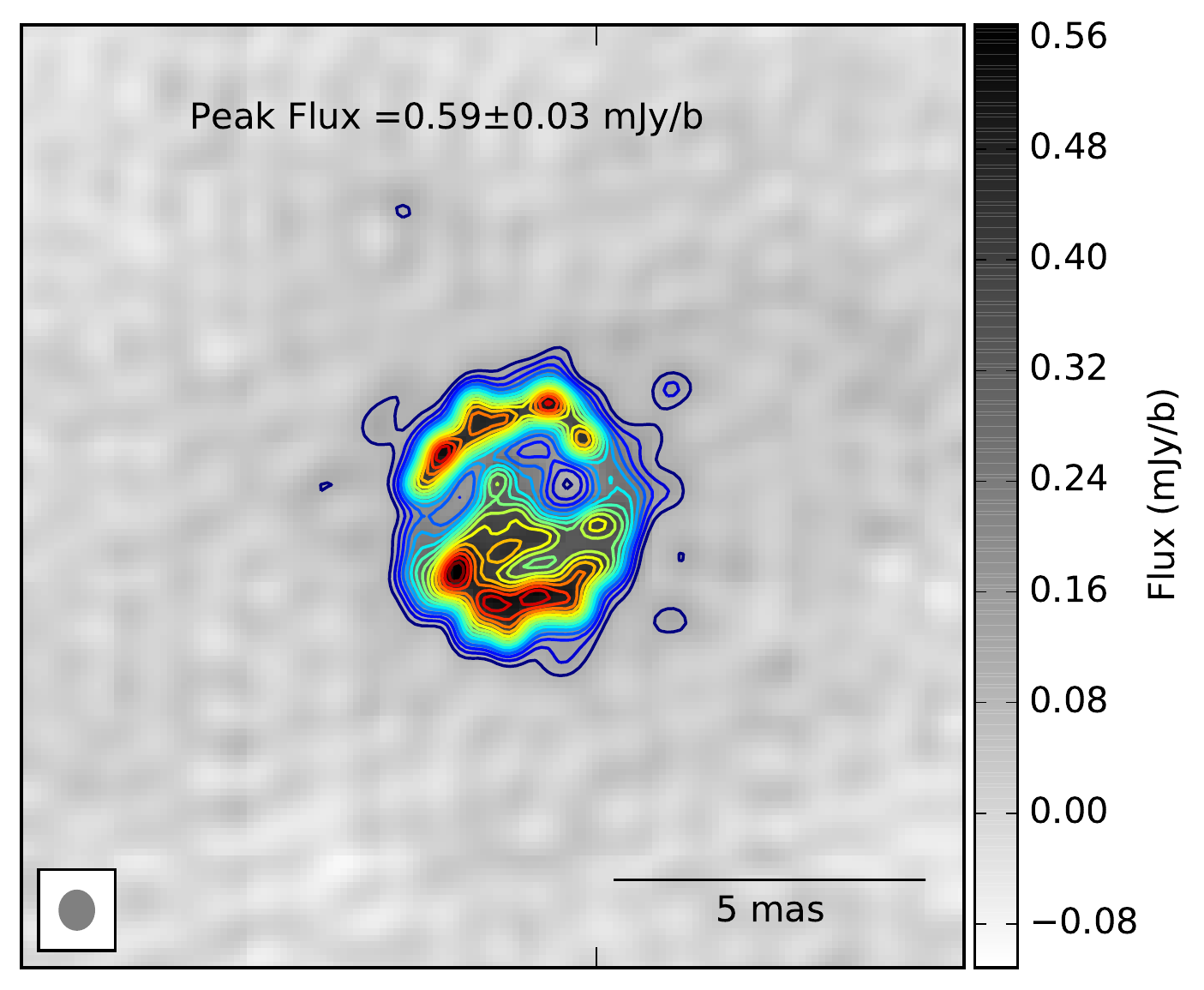}
    \label{fig:test03}}
\end{minipage}%
\begin{minipage}{0.25\linewidth}
  \centering
  \subcaptionbox{Day 807}
  {\includegraphics[width=\linewidth]{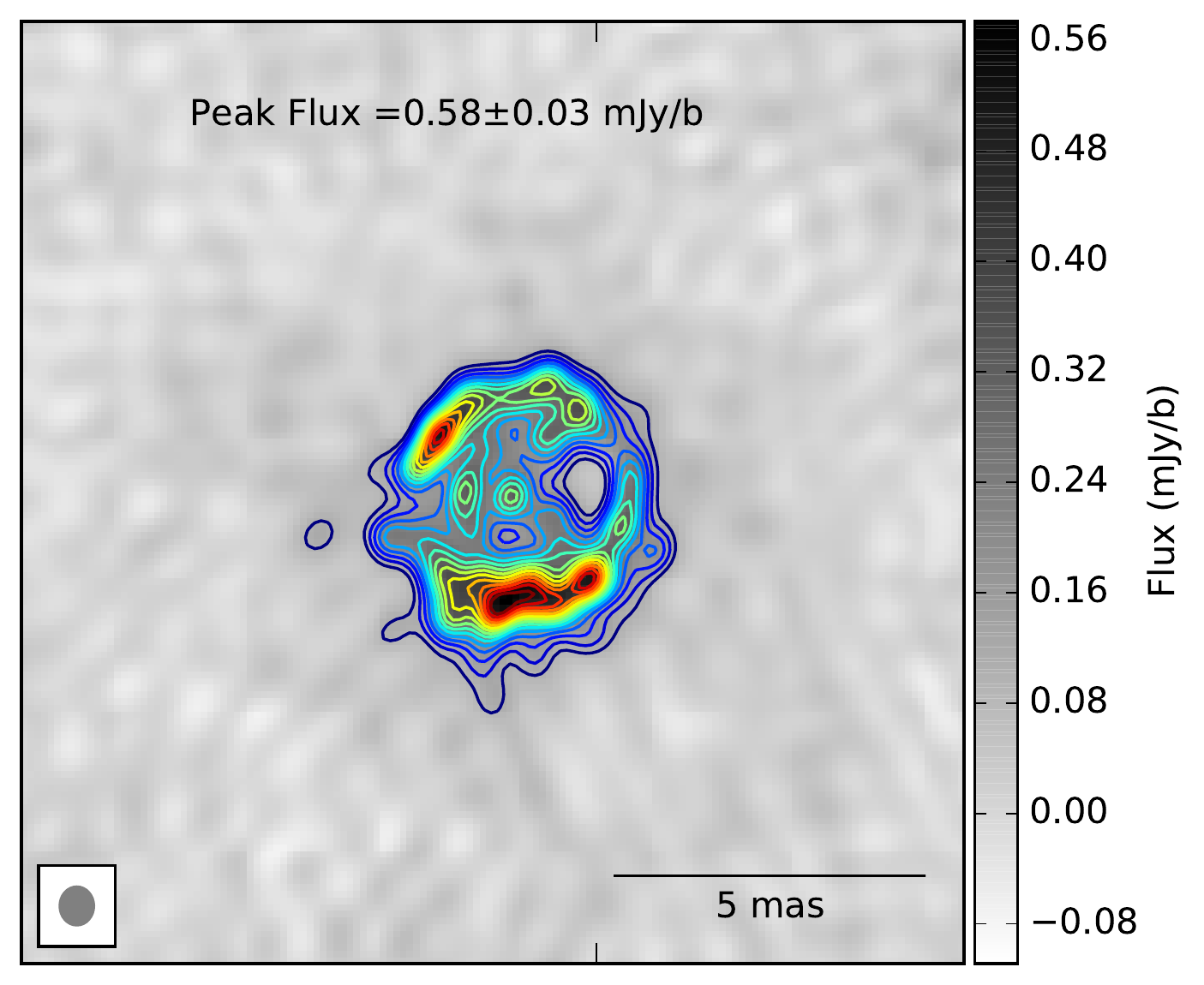}
    \label{fig:test04}}
\end{minipage}%

\begin{minipage}{0.25\linewidth}
  \centering
  \subcaptionbox{Day 866}
  {\includegraphics[width=\linewidth]{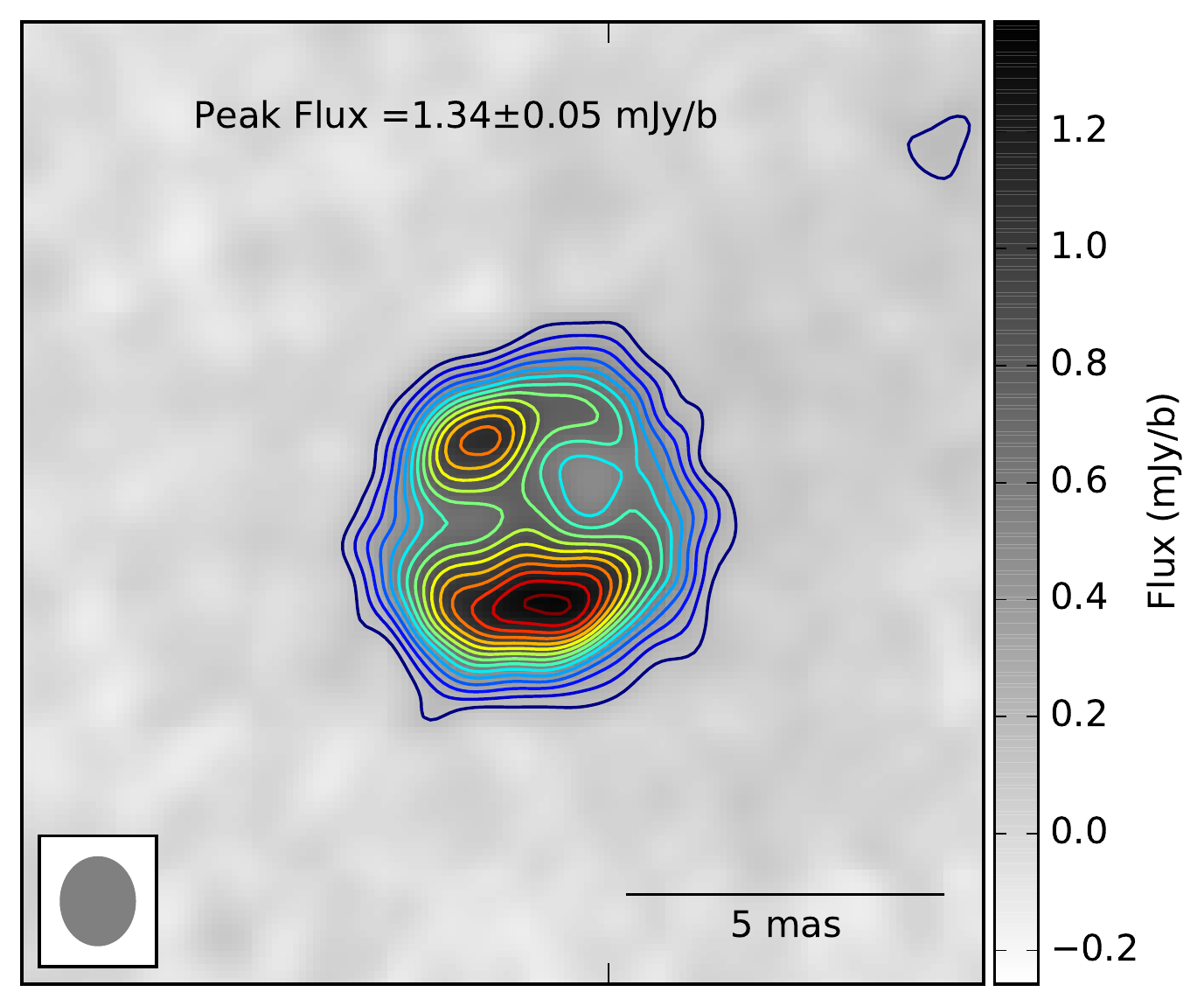}
    \label{fig:test05}}
\end{minipage}%
\begin{minipage}{0.25\linewidth}
  \centering
  \subcaptionbox{Day 931}
  {\includegraphics[width=\linewidth]{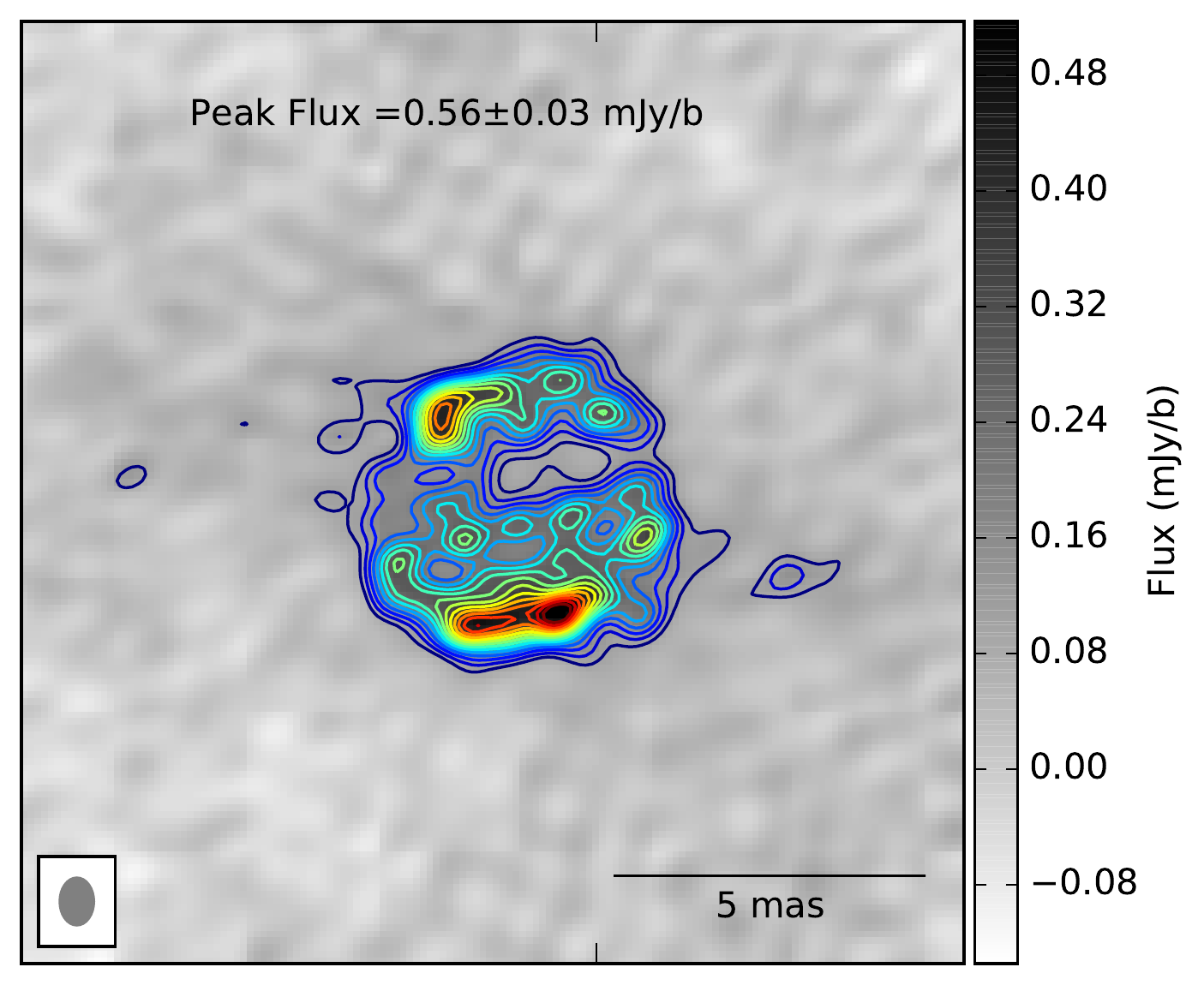}
    \label{fig:test06}}
\end{minipage}%
\begin{minipage}{0.25\linewidth}
  \centering
  \subcaptionbox{Day 1006}
  {\includegraphics[width=\linewidth]{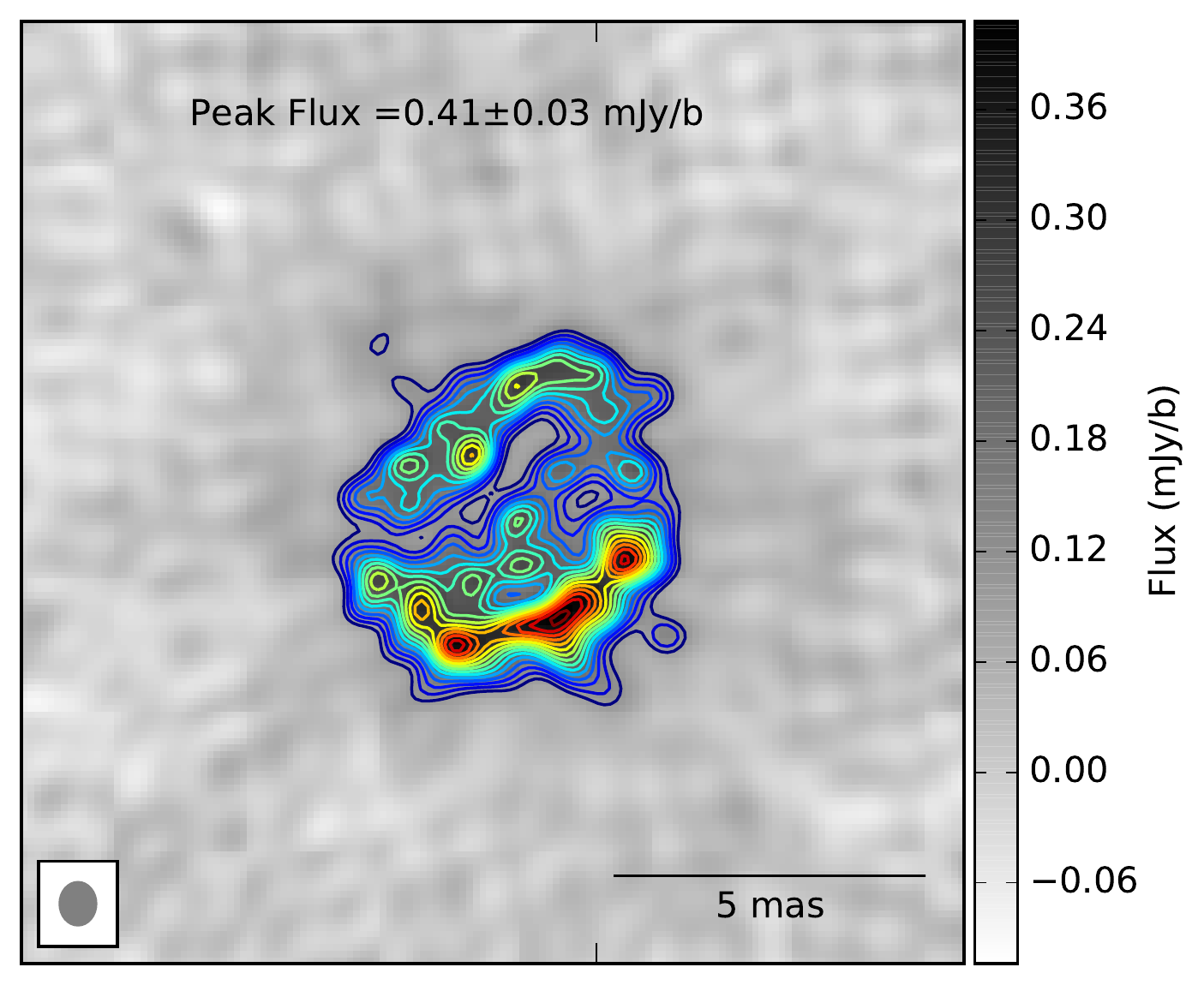}
    \label{fig:test07}}
\end{minipage}%
\begin{minipage}{0.25\linewidth}
  \centering
  \subcaptionbox{Day 1055}
  {\includegraphics[width=\linewidth]{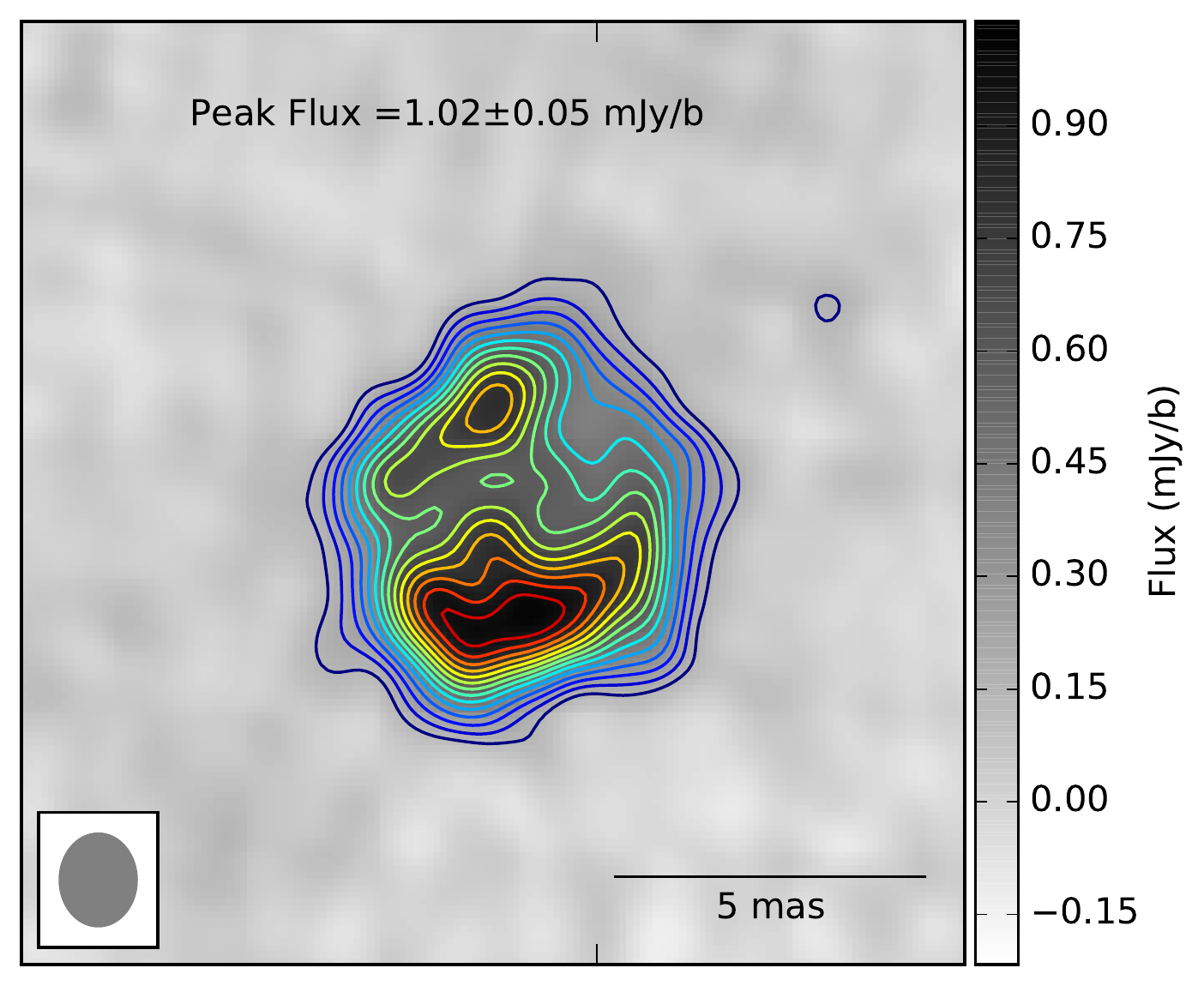}
    \label{fig:test08}}
\end{minipage}%

\begin{minipage}{0.25\linewidth}
  \centering
  \subcaptionbox{Day 1125}
  {\includegraphics[width=\linewidth]{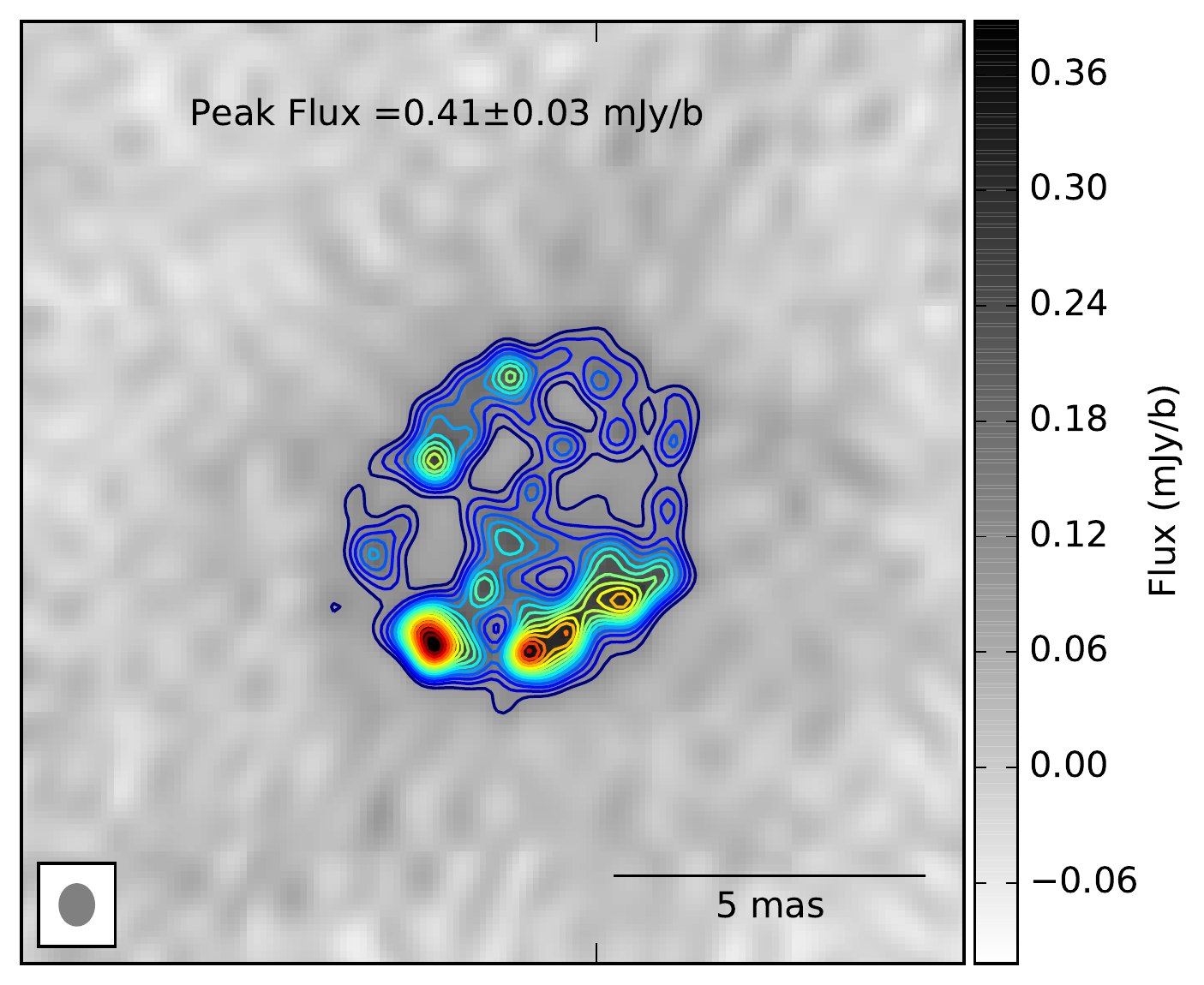}
    \label{fig:test09}}
\end{minipage}%
\begin{minipage}{0.25\linewidth}
  \centering
  \subcaptionbox{Day 1180}
  {\includegraphics[width=\linewidth]{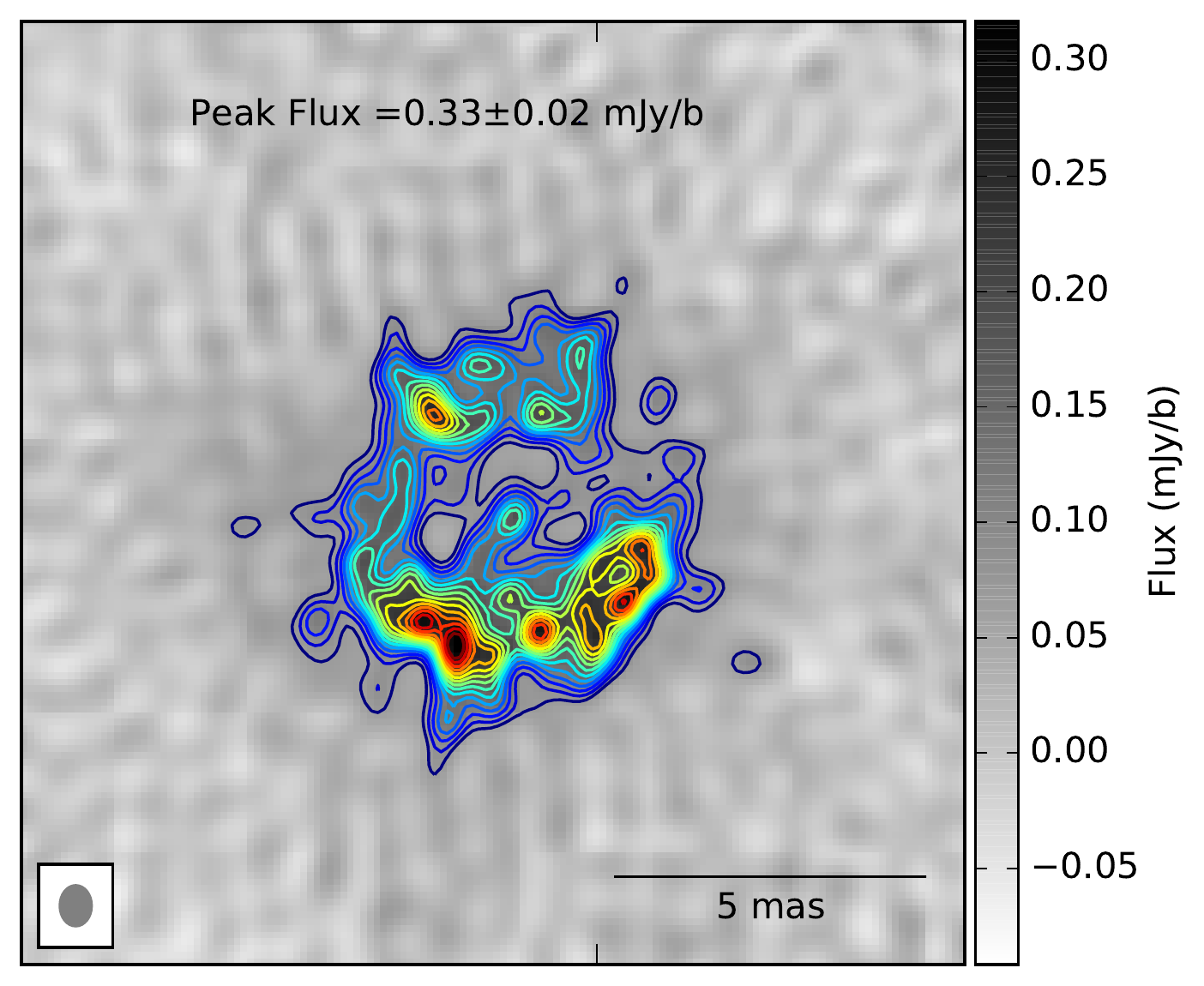}
    \label{fig:test010}}
\end{minipage}%
\begin{minipage}{0.25\linewidth}
  \centering
  \subcaptionbox{Day 1274}
  {\includegraphics[width=\linewidth]{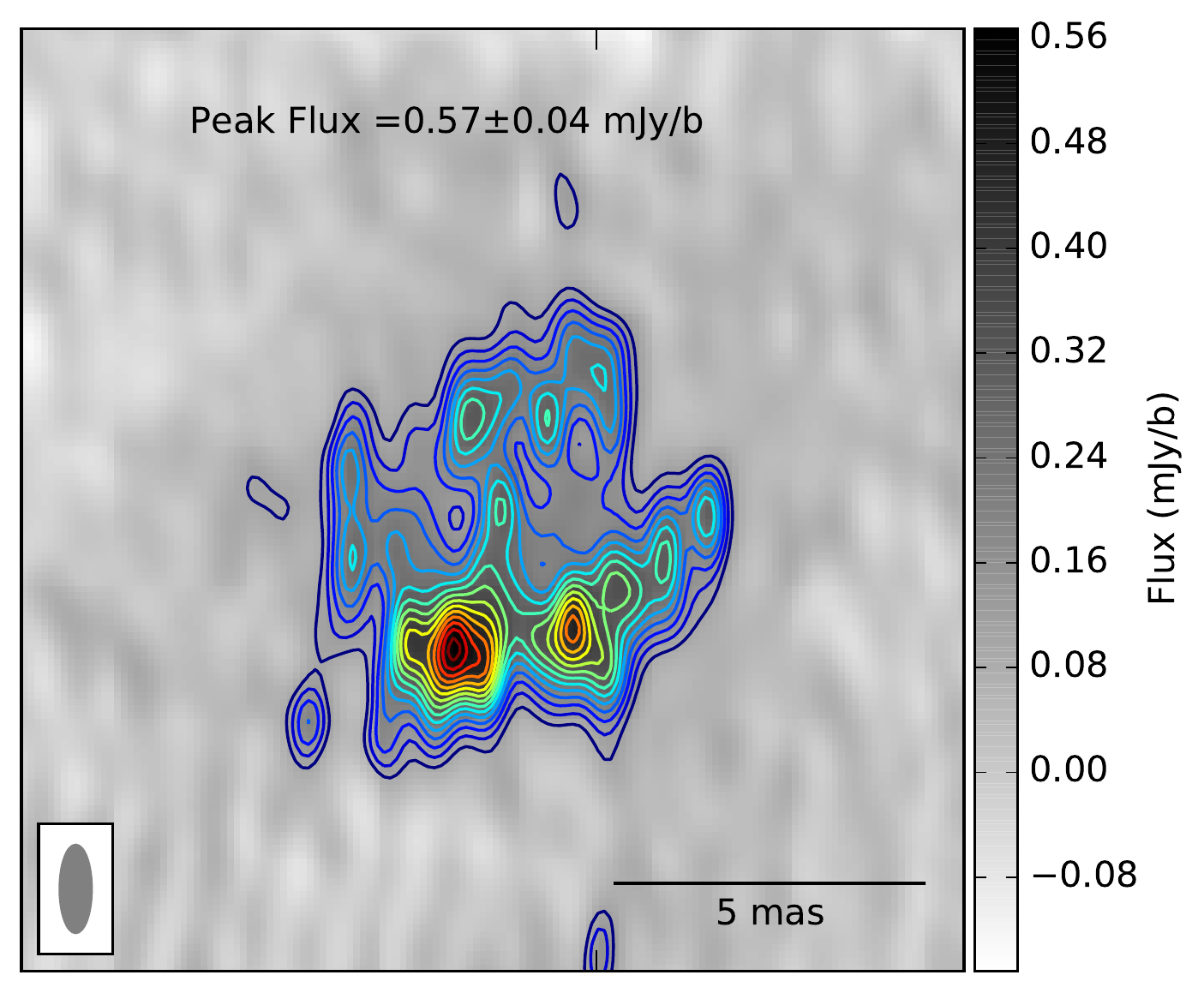}
    \label{fig:test011}}
\end{minipage}%
\begin{minipage}{0.25\linewidth}
  \centering
  \subcaptionbox{Day 1314}
  {\includegraphics[width=\linewidth]{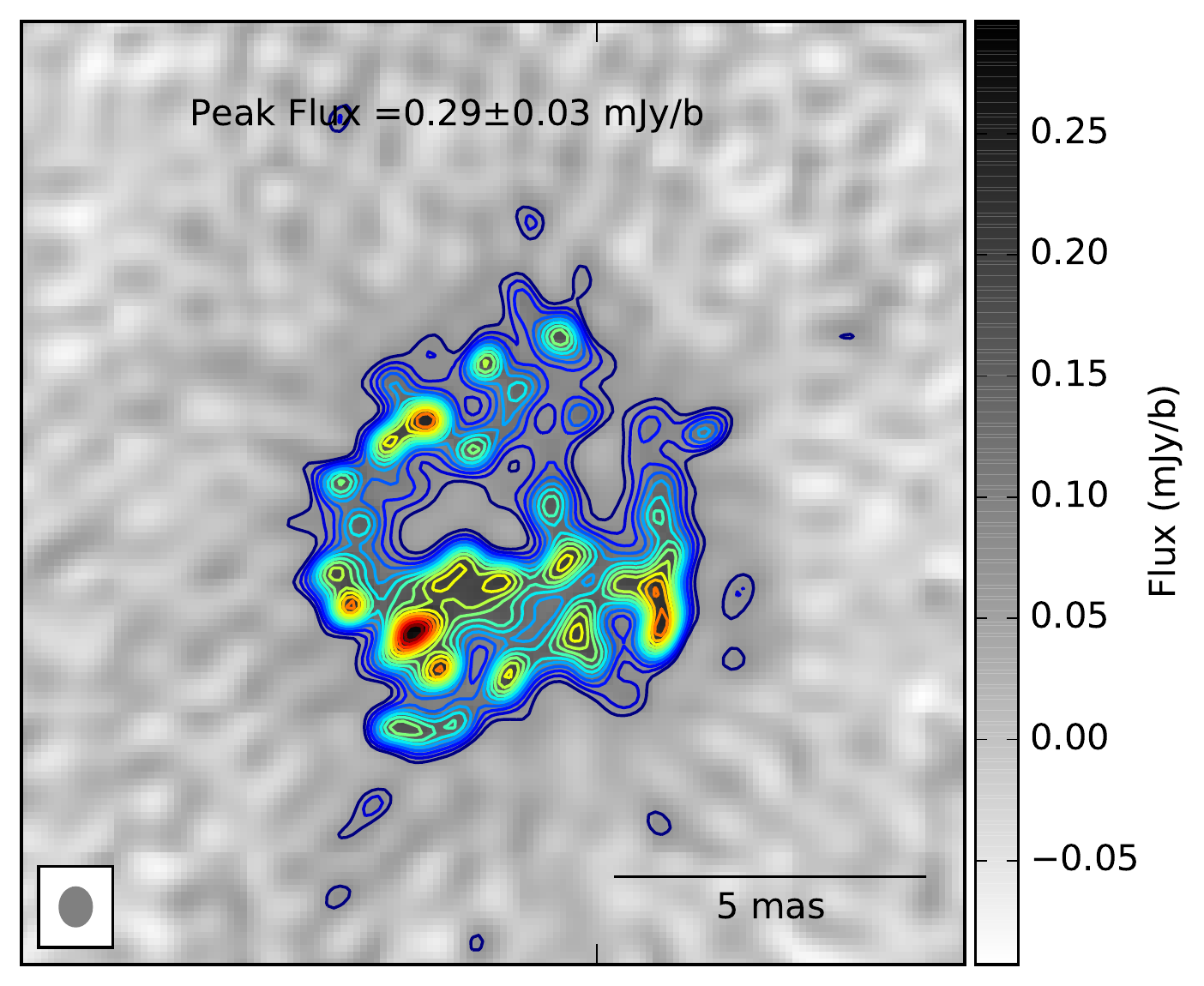}
    \label{fig:test012}}
\end{minipage}%

\begin{flushleft}
\begin{minipage}{0.25\linewidth}
  \centering
  \subcaptionbox{Day 1401}
  {\includegraphics[width=\linewidth]{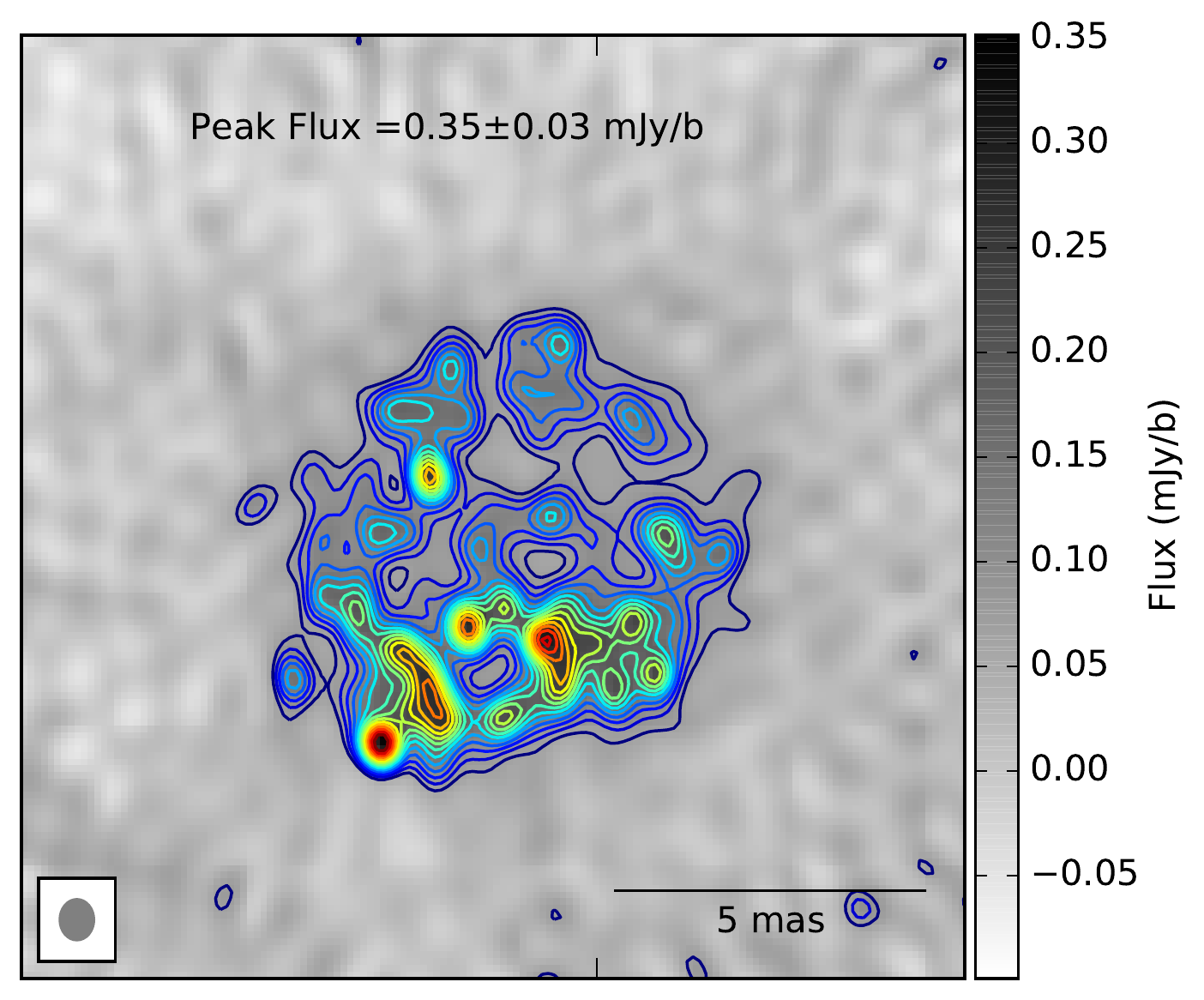}
    \label{fig:test013}}
\end{minipage}%
\begin{minipage}{0.25\linewidth}
  \centering
  \subcaptionbox{Day 1650}
  {\includegraphics[width=\linewidth]{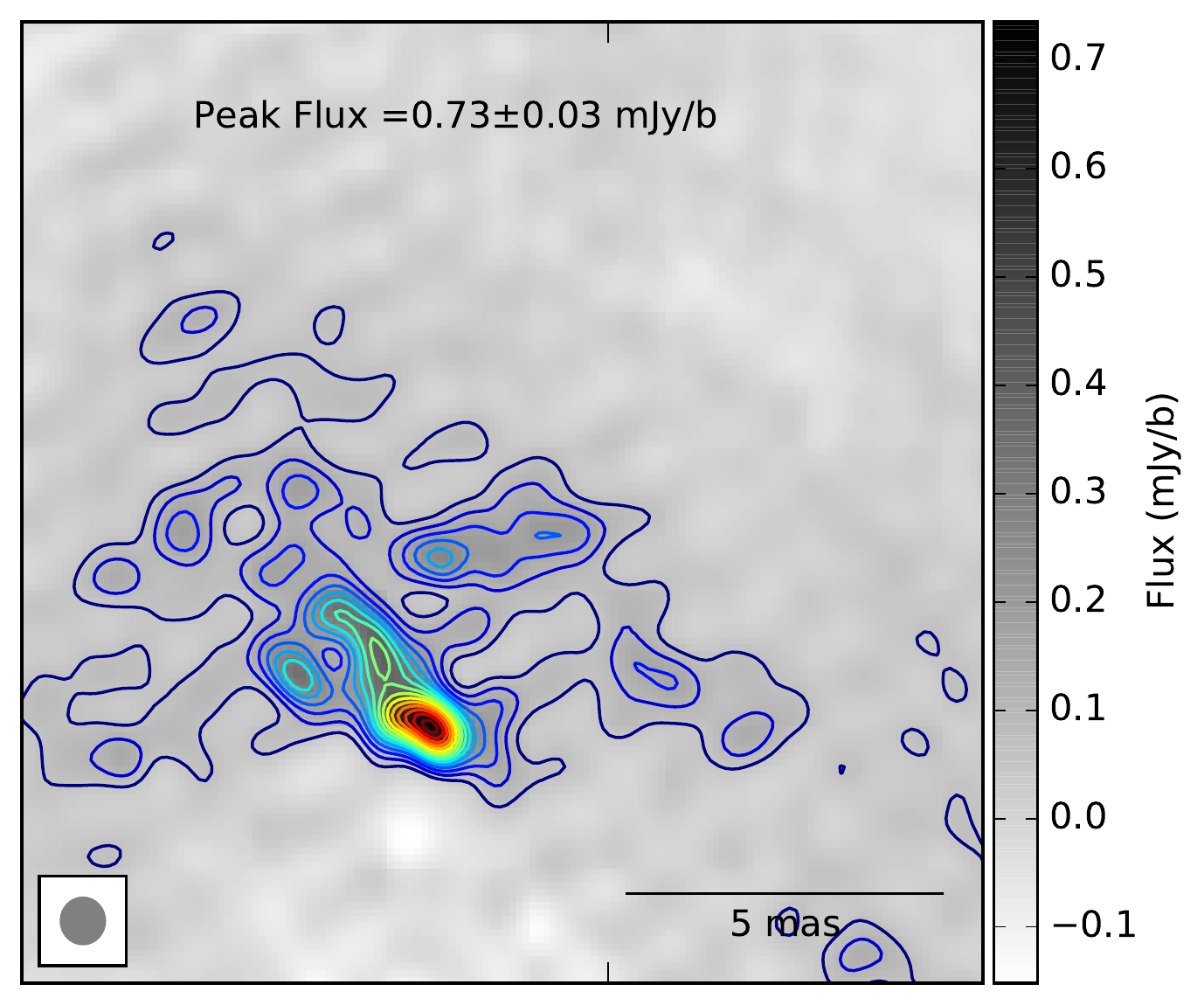}
    \label{fig:test014}}
\end{minipage}%
\begin{minipage}{0.25\linewidth}
  \centering
  \subcaptionbox{Day 1806}
  {\includegraphics[width=\linewidth]{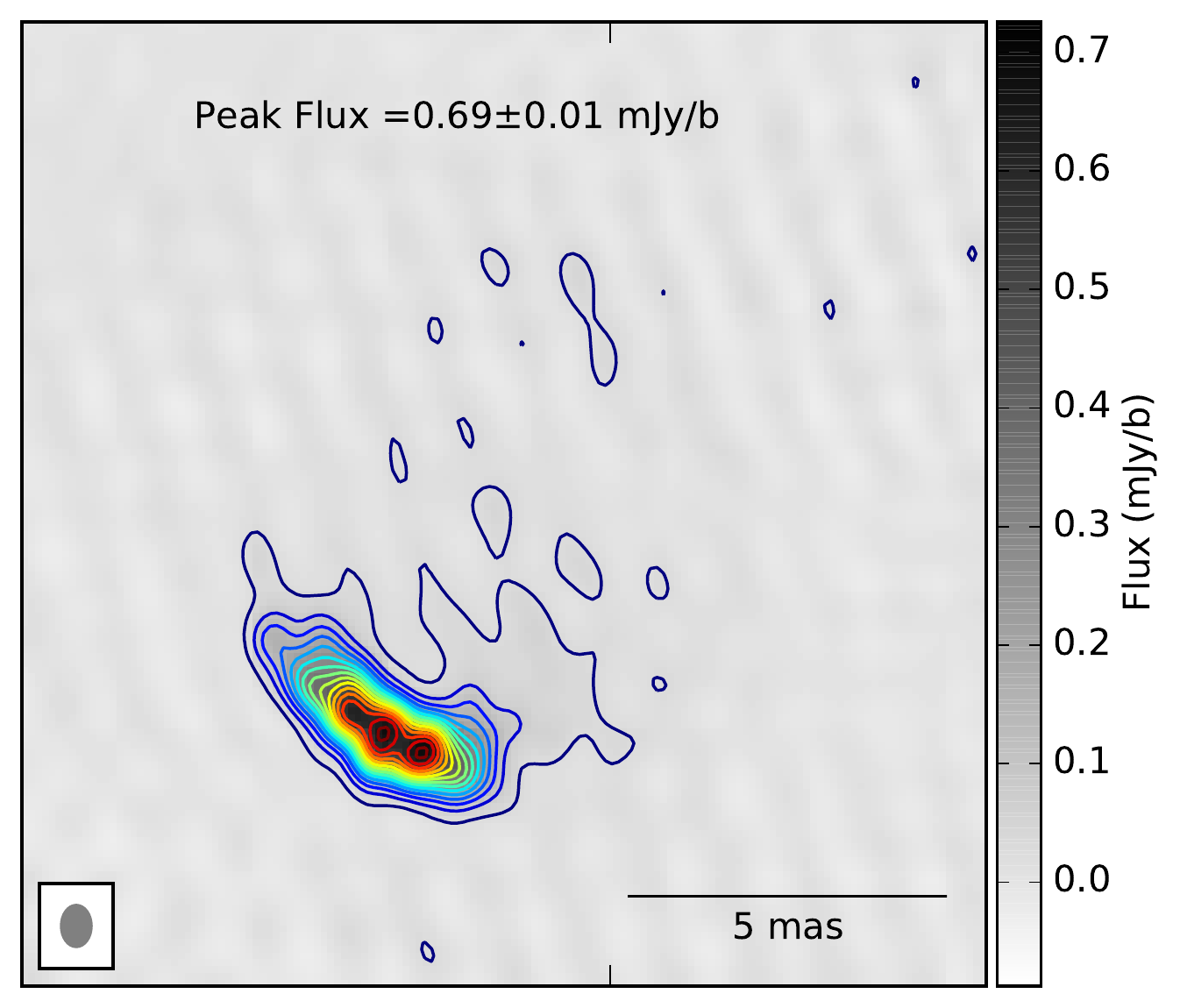}
    \label{fig:test015}}
\end{minipage}%
\end{flushleft}
\caption{The natural weighted VLBI images of the expanding shell produced by
supernova SN$\,$2008iz at 8.4$\,$GHz. They cover the observation periods between 2009 October and 2013 January as summarized in Table$\,$\ref{table1b}.
The figures show the 8.4$\,$GHz linear gray scale images overlaid
with their contour maps that scale to the peak intensity at each epoch
(listed in each panel). All image contours correspond to 6\% of the
peak intensity (lowest contour) and subsequent increases by 6\%. The
images have been convolved with the natural weighting beam, shown at the bottom left of each panel.}
  \label{vlbi_xband}
\end{figure*}

The VLBI images of SN$\,$2008iz at 4.8 and 8.4$\,$GHz are shown
in Fig.\ref{vlbi_cband}, and \ref{vlbi_xband}
respectively. All the images are phase referenced to the M$\,$81 core, M$\,$81*, for consistency in their alignment. The SNR is resolved at these frequencies. We use natural weighting in order to reveal more information on the larger
scale structures. The size and shape of the beam used for restoring
the images is determined by the set of antennas (uv-coverage) that were employed
for each of the the considered data sets. For instance, the resolution
of images Fig \ref{vlbi_cband}$\,$e,$\,$f and Fig \ref{vlbi_xband}$\,$e,$\,$h
is lower because some telescopes, especially the Effelsberg 100 meter
telescope, have not taken part in the observations.

The images show that the supernova expansion starts with a shell like
structure that gets more and more asymmetric. Overall, we see expansion
with similar velocities into all directions. During the later epochs,
the emission distribution breaks up and shows substructures in the
southern part of the ring. There, the ring is much brighter than in
its other parts, indicating a denser surrounding medium along the
southern direction. 

The deceleration of SN$\,$2008iz is derived from the VLBI images convolved
with a dynamic beam at 4.8 and 8.4$\,$GHz. A dynamic beam (see \citealt{Marcaide1997}) scales the resolution of the later epochs to the resolution of the first epoch with respect to the size of the supernova shell size before imaging. It eliminates the systematic errors that would otherwise arise from the different resolution with which the SNR was imaged while it expands. We obtain the size of the SNR with the AIPS task 'IRING', which was used to fit 2D Gaussians to the obtained light profiles. The centre of the SNR remains at the same position during the time span of our observations. We take the radius of the supernova ring at 50\% of the peak intensity on the outside of the bright rim since this is the real ring size times an unknown factor. This method is more reliable than taking the maximum of the ring as the radius (see a more detailed description in \citealt{Brunthaler2010, Beswick2006}). We fit a power-law of the form 

\begin{equation}
\mathrm{R_{50\%}=c_{y}(t-t_{0})^{m}}
\end{equation}

\noindent to the size evolution of SN$\,$2008iz, where c$_{y}$ is a scaling
factor, (t-t$_{0}$) is the time elapsed after the explosion date
and $\mathit{m}$ is the deceleration index. We derive values of c$_{y}$=(7.5$\pm$0.9)$\times$
10$^{-6}$ arcsec/day and $\mathit{m}$=0.86$\pm$0.02. The errors
are obtained from the post-fit covariance matrix. The size evolution
and deceleration of SN$\,$2008iz are displayed in Fig. \ref{deceleration}.
Also shown are two 22$\,$GHz data points from \citet{Brunthaler2010}
that were included in the analysis. The last two data points were
excluded from the fit since we know from the light curve that there
is a change in the surrounding medium of SN$\,$2008iz at around 1400
days after the explosion that is not caused by internal processes.

\begin{figure}[h!]
  \resizebox{0.9\hsize}{!}{\includegraphics{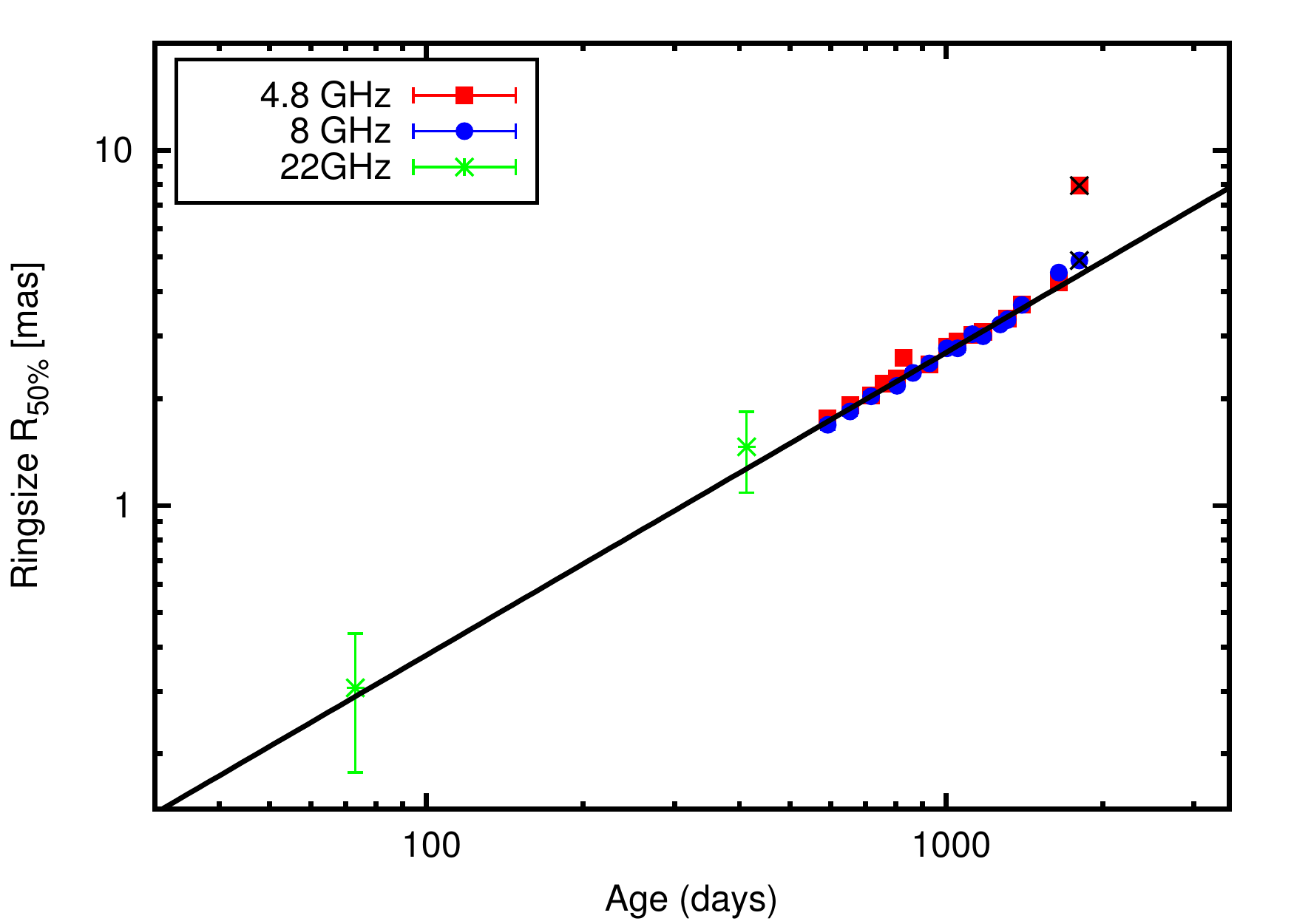}}
  \caption{The expansion curve of SN$\,$2008iz. The two data points at 22$\,$GHz
were taken from \citet{Brunthaler2010}, while the last two data points
(overlaid with a black `$\mathrm{x}$' symbol) were excluded from the fit.}
  \label{deceleration}
\end{figure}

\begin{table*}
\caption{The log of SN\,2008iz integrated radio flux measurements observed with VLA.}
\label{table1}
\centering
\begin{tabular}{c c c c c c c c c }
\hline
 Date & Days since &Config- &S$_{1.4\,\mathrm{GHz}}$ & S$_{4.8\,\mathrm{GHz}}$ & S$_{8.4\,\mathrm{GHz}}$ & S$_{22.3\,\mathrm{GHz}}$& S$_{35\,\mathrm{GHz}}$ & S$_{43.2\,\mathrm{GHz}}$ \\
  (dd/mm/yy) & 18-2-2008 &uration &(mJy)  & (mJy) & (mJy) & (mJy)& (mJy) & (mJy) \\ 
\hline 
\hline
2008/03/24 & 36 & C & -- & -- & -- & 100$\pm$2 & -- & --\\
2008/05/03 & 76 & C & -- & -- & -- & 88.4$\pm$0.2 & -- & --\\
2009/04/08 & 416 & B & -- & -- & -- & 9.2$\pm$0.2 & -- & --\\
2009/04/27 & 435 & B & 55$\pm$6 & 30$\pm$5 & 14$\pm$3 & 6.2$\pm$0.7 & -- & 2.5$\pm$0.4\\
2009/07/24 & 523 & C & -- & 21$\pm$2 & 11$\pm$1 & -- & -- & 1.0$\pm$0.6\\
2009/09/19 & 580 & B & -- & 14$\pm$2 & 6$\pm$2 & 5$\pm$2 &  & 1.8$\pm$0.5\\
2009/10/21 & 612 & D & -- & -- & -- & 3$\pm$1 & -- & --\\
2010/10/20 & 976 & C & -- & -- & -- & 2.4$\pm$0.3 & -- & 1.1$\pm$0.2\\
2010/12/18 & 1035 & C & -- & -- & 10.1$\pm$0.9 & 2.8$\pm$0.3 & -- & 1.7$\pm$0.4\\
2011/02/12 & 1091 & B & -- & 10.2$\pm$0.6 & 7.7$\pm$0.7 & 3.5$\pm$0.1 & -- & 0.9$\pm$0.2\\
2011/05/21 & 1189 & AB & 23$\pm$5 & 6$\pm$2 & 6.4$\pm$0.7 & 3.1$\pm$0.3 & -- & 0.6$\pm$0.1\\
2011/07/30 & 1259 & A & 19$\pm$3 & 12.3$\pm$0.9 & 7.5$\pm$0.5 & -- & -- & --\\
2011/08/01 & 1261 & A & 17$\pm$2 & 11$\pm$1 & 7.1$\pm$0.4 & -- & -- & --\\
2011/08/09 & 1269 & A & -- & -- & -- & 2.3$\pm$0.2 & -- & 0.2$\pm$0.1\\
2012/02/02 & 1446 & C & -- & -- & -- & 1.9$\pm$0.5 & 0.9$\pm$0.2 & --\\
2012/08/25 & 1651 & B & -- & 12$\pm$2 & 9$\pm$2 & 3.3$\pm$0.6 & 2.0$\pm$0.2 & --\\
2012/11/21 & 1739 & A & -- & 26$\pm$1 & 16.8$\pm$0.9 & 5.0$\pm$0.2 & 2.7$\pm$0.2 & --\\
2014/01/23 & 2167 & AB & -- & 41$\pm$7 & -- & 6.5$\pm$0.6 & -- & --\\
\hline
\end{tabular} 
\tablefoot{
The data on days 24 Mar 2008 to 27 Apr 2009 are obtained from \cite{Brunthaler2009b} and \cite{Brunthaler2010}.
}
\end{table*}

\begin{table*}
\caption{The log of SN$\,$2008iz radio flux and angular size measurements
obtained from the VLBI observations.}
\label{table1b}
\centering
\begin{tabular}{cccccccc}
\hline 
Date & Age & S$_{1.6GHz}$ & S$_{1.6GHz}$ & S$_{4.8GHz}$ & S$_{8.4GHz}$ & $\Theta_{R_{50\%}\,4.8GHz}$ & $\Theta_{R_{50\%}\,8.4GHz}$\tabularnewline
 & (t-t$_{0}$) & (mJy) & (mJy/b) & (mJy/b) & (mJy/b) & (mas) & (mas)\\
\hline 
2009/10/04 & 595 & 64$\pm$6 & 57.9$\pm$0.5 & 3.02$\pm$0.04 & 0.83$\pm$0.04 & 1.8$\pm$0.2 & 1.7$\pm$0.2\\
2009/12/04 & 656 & 55$\pm$4 & 50.9$\pm$0.4 & 2.61$\pm$0.03 & 0.78$\pm$0.03 & 1.9$\pm$0.2 & 1.8$\pm$0.1\\
2010/02/06 & 720 & 42$\pm$8 & 34.6$\pm$0.2 & 2.09$\pm$0.03 & 0.59$\pm$0.03 & 2.0$\pm$0.2 & 2.0$\pm$0.1\\
2010/03/20 & 758 & -- & -- & 1.80$\pm$ 0.03 & -- & 2.3$\pm$0.2 & --\\
2010/05/04 & 807 & 39$\pm$9 & 30.8$\pm$0.2 & 1.79$\pm$0.03 & 0.58$\pm$0.03 & 2.3$\pm$0.2 & 2.2$\pm$0.1\\
2010/05/29 & 828 & -- & -- & 1.71$\pm$0.03 & -- & 2.4$\pm$0.2 & --\\
2010/06/07 & 840 & 32$\pm$8 & 24.9$\pm$0.1 & -- & -- & -- & --\\
2010/07/02 & 866 & 35$\pm$8 & 26.5$\pm$0.2 & 4.56$\pm$0.05 & 1.34$\pm$0.05 & -- & 2.4$\pm$0.2\\
2010/09/05 & 931 & 30$\pm$9 & 20.8$\pm$0.1 & 3.73$\pm$0.04 & 0.56$\pm$0.03 & 2.5$\pm$0.2 & 2.5$\pm$0.2\\
2010/11/19 & 1006 & 30$\pm$9 & 21.5$\pm$0.1 & 1.73$\pm$0.03 & 0.41$\pm$0.03 & 2.8$\pm$0.2 & 2.8$\pm$0.2\\
2011/01/07 & 1055 & 27$\pm$9 & 17.6$\pm$0.1 & 1.15$\pm$0.03 & 1.02$\pm$0.05 & 2.9$\pm$0.3 & 2.8$\pm$0.2\\
2011/03/18 & 1125 & 26$\pm$9 & 17.1$\pm$0.1 & 0.98$\pm$0.03 & 0.41$\pm$0.03 & 3.0$\pm$0.3 & 3.0$\pm$0.2\\
2011/05/12 & 1180 & 26$\pm$9 & 17.3$\pm$0.2 & 0.76$\pm$0.03 & 0.33$\pm$0.02 & 3.1$\pm$0.3 & 3.0$\pm$0.2\\
2011/08/14 & 1274 & 25$\pm$10 & 15.2$\pm$0.1 & 1.55$\pm$0.05 & 0.57$\pm$0.04 & -- & 3.2$\pm$0.2\\
2011/08/14 & 1274 & 24$\pm$9 & 15.3$\pm$0.1 & 3.78$\pm$0.04 & 1.18$\pm$0.04 & -- & --\\
2011/09/23 & 1314 & 25$\pm$2 & 12.6$\pm$0.1 & 0.91$\pm$0.03 & 0.29$\pm$0.03 & 3.4$\pm$0.3 & 3.3$\pm$0.2\\
2011/09/23 & 1314 & 21$\pm$9 & 12.0$\pm$0.1 & 3.17$\pm$0.07 & 0.90$\pm$0.03 & -- & 3.3$\pm$0.2\\
2011/12/19 & 1401 & 28$\pm$9 & 18.7$\pm$0.2 & 0.75$\pm$0.03 & 0.35$\pm$0.03 & 3.7$\pm$0.6 & 3.7$\pm$0.3\\
2012/08/25 & 1650 & 39$\pm$9 & 30.3$\pm$0.3 & 2.43$\pm$0.05 & 0.73$\pm$0.03 & 4.3$\pm$0.4 & 4.5$\pm$0.2\\
2013/01/28 & 1806 & 40$\pm$9 & 21.2$\pm$0.1 & 1.65$\pm$0.01 & 0.69$\pm$0.01 & 8.0$\pm$0.4 & 4.9$\pm$0.2\\
\hline 
\end{tabular} 
\tablefoot{
Both peak and integrated flux density values of our 1.6$\,$GHz are displayed. The peak flux values are presented for the 4.8$\,$GHz and 8.4$\,$GHz measurements because the supernova SN$\,$2008iz was resolved at these frequencies. The $\mathrm{\Theta}_{R_{50\%}}$ are the angular sizes derived as the radius of the supernova ring at 50\% of the peak intensity on the outside of the bright rim.}
\end{table*}

\clearpage

\clearpage

\section{Discussion}
\subsection{Flux-density flare from day $\sim$970}
\begin{figure}[h!]
  \resizebox{0.9\hsize}{!}{\includegraphics{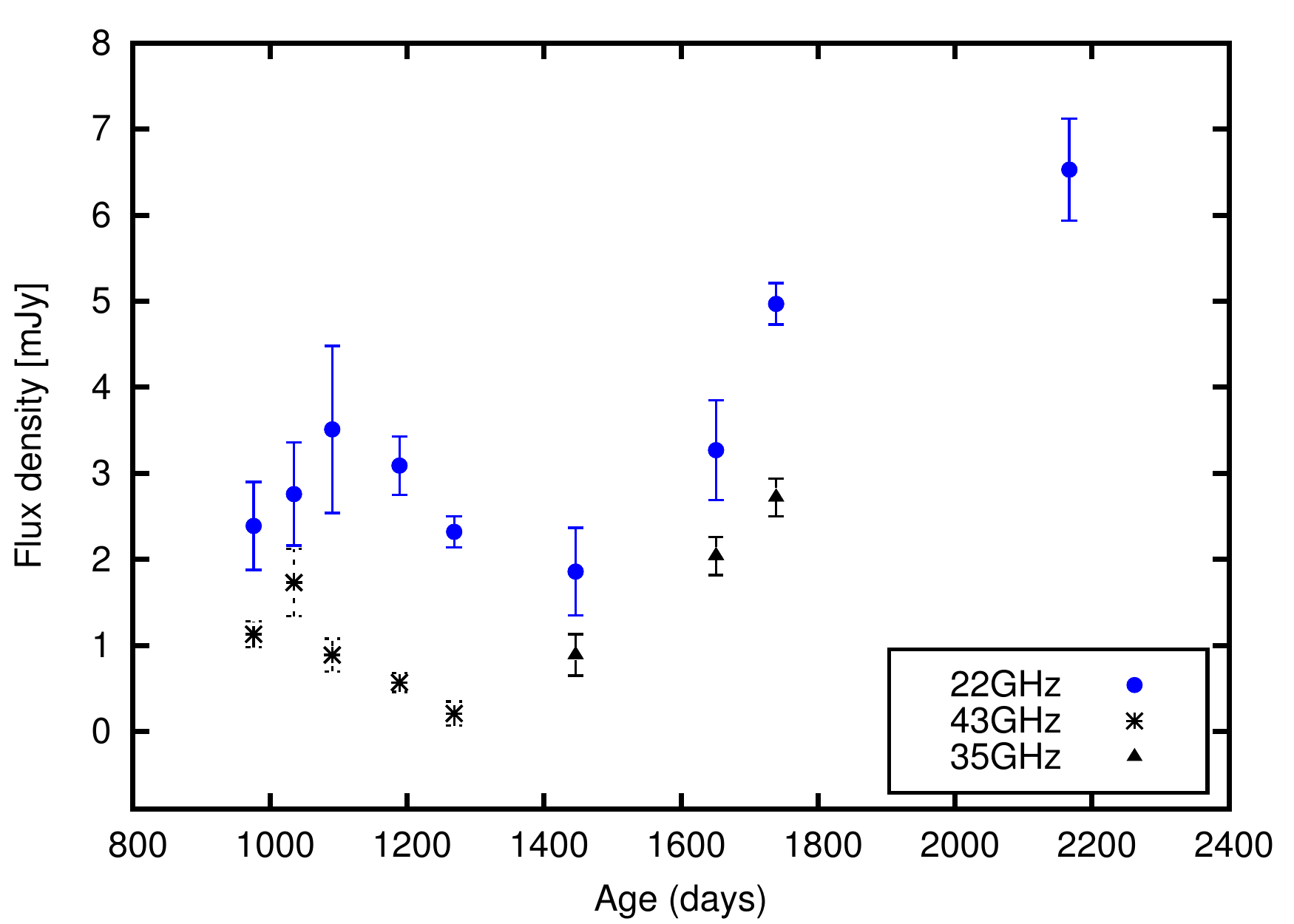}}
  \caption{A zoom into our higher frequencies light curve for a period between t=\,800 to t=2400\,days. The flux density enhancement at t\,$\ge$\,1400 days was detected in both VLA and VLBI results at all frequencies.}
  \label{bump}
\end{figure}

A zoom into the light curve at the higher frequencies (see Fig.\ref{bump})
uncovers excess emission from the supernova after t=970 days and after
t=1400 days. We note that the extra emission after t=970 days is detected
at 8.4, 22.3 and 43.2$\,$GHz while the enhancement at t=1400 days
is detected at 4.8, 8.4, 22 and 35$\,$GHz. The later flare after
time $\geq$$\,$1400 days is also confirmed with our densely sampled
VLBI 1.6$\,$GHz results. The extra emission increases the total flux
density of the supernova by a factor of $\sim$$\,$2 at t=970 days,
lasting for about 240 days. The second flare increases the total flux
density by a factor of $\sim$4, without showing signs of flux decline
so far, i.e., in 2014 January. This calls for a continuation of the
monitoring to determine how long the second flare lasts. 

The source is obviously optically-thin at all frequencies during the
flares and its spectral index remains steep (see section \ref{spectralindex}).
\citet{Weiler2002} indicate that a supernova whose radio emission
preserves its spectral index while deviating from the standard model
is showing evidence for a change in the conditions of the CSM that
its remnant expands into. We therefore examine the changes in the
density of the CSM as the explanation to the observed flares. According
to \citet{Chevalier1982}, radio luminosity is related to the average
CSM density ($\rho_{csm}$$\propto$ $\dot{M}$/w) through 

\begin{equation}
L\propto\left(\frac{\dot{M}}{w}\right)^{(\gamma-7+12m)/4},
\label{eq:chevalia82}
\end{equation}

\noindent where the relativistic particle index $\gamma$=1$-$2$\alpha$, $\alpha$=$-$1.08$\pm$0.08
and expansion index $\mathit{m}$= 0.86$\pm$0.02. For SN$\,$2008iz,
L$\propto$($\dot{M}$/w)$^{1.62}$. Consequently, for the radio emission
to double its flux in the first flare, the CSM density must have increased
by a factor of 1.5, while for the increase by factor of 4 in the second
flare, the CSM density must have tripled. However, the VLBI images
of the remnant of SN$\,$2008iz do not show a spherically-symmetric
density enhancement. This suggests that the shock wave is encountering
dense inhomogeneities in the CSM.

The late flare (day 1400), may also be related to the transition of
the front shock from the CSM bubble into the ISM. This would enhance
the magnetic field and increase the emission at all frequencies. However,
we obtain the size of the shock on day 1400 is $\sim$0.06$\,$pc
(3.6$\,$mas), which is very small compared to the supernova remnant
sizes measured by \citet{Batejat2011} of $\sim$0.4$\,$pc in Arp$\,$220
at the transition. Were the flare related to the transition, this
indicates the presence of a very dense ISM for the pressure balance
to hold at the bubble boundary of SN$\,$2008iz. Thus we require an
ISM with a density higher than the characteristic one in Arp$\,$220
of 10$^{4}$cm$^{-3}$ to explain the light curve of SN$\,$2008iz
at late stages, in the frame of the ISM interaction model.

\subsection{Evolution of the spectral index}\label{spectralindex}

\begin{figure}[h!]
  \resizebox{0.9\hsize}{!}{\includegraphics{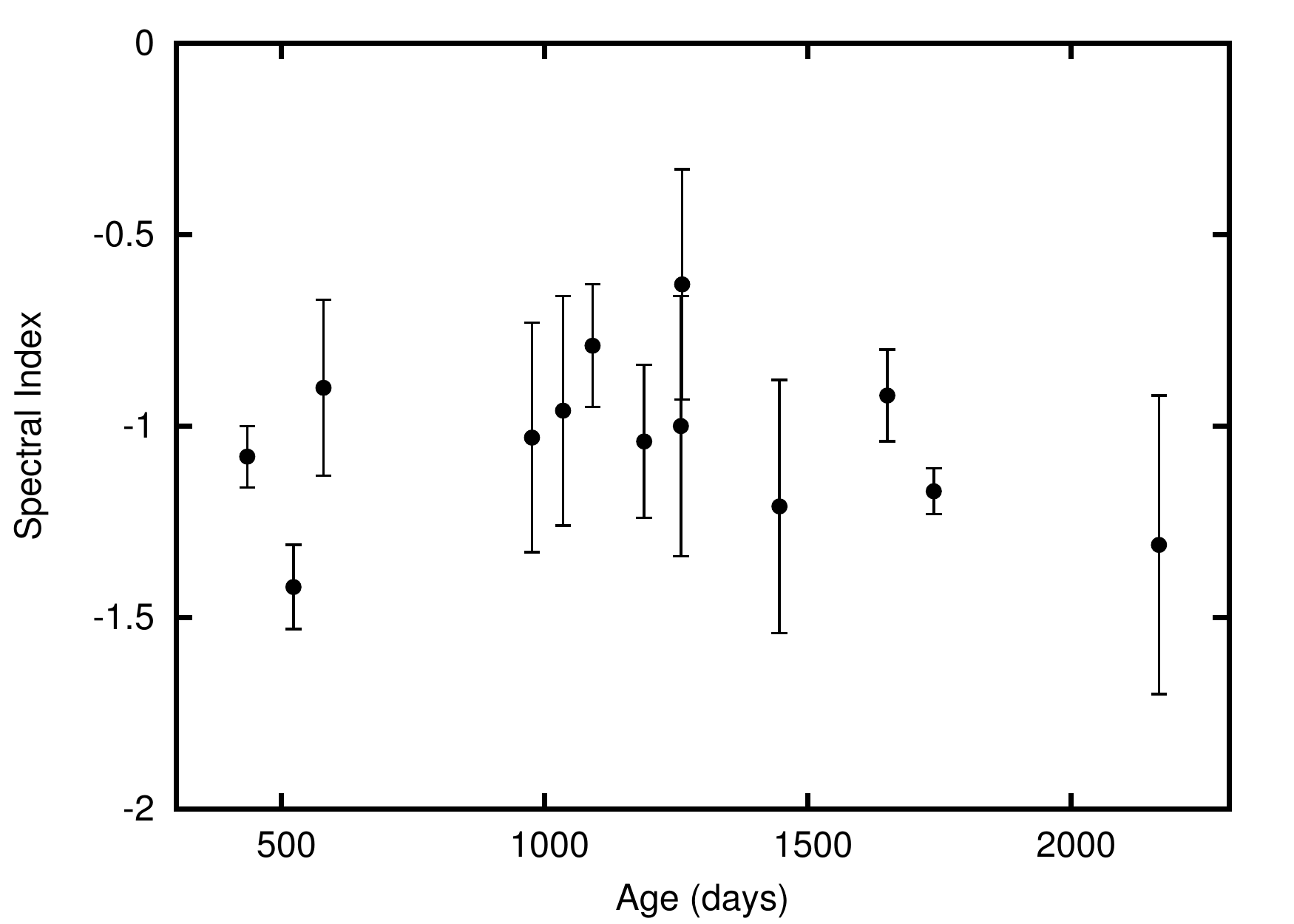}}
  \caption{The spectral index, $\alpha$, for SN$\,$2008iz obtained from fitting
a simple power law fit to the radio continuum data at each epoch.
The spectral index remains steep, near $\alpha=$$-$1.}
  \label{index}
\end{figure}

Figure \ref{index} shows the best-fit radio spectral
index, $\alpha$, for SN$\,$2008iz from $\sim$430 days up to 2167
days. To obtain the radio spectral index we fit a simple power-law
spectrum of the form ($S=S_{0}\nu^{\alpha}$) to our VLA data for
all the epochs except for epoch 2009 April 27 which was best fit with
a broken power-law of the form 

\begin{equation}
S=S_{0}\left(\frac{\nu}{\nu_{0}}\right)^{\alpha}\left(1-e^{-\left(\frac{\nu}{\nu_{0}}\right)^{\delta-\alpha}}\right)
\end{equation}
as presented by \citet{Brunthaler2010}. We fit this expression to
our data with the S$_{0}$ and $\alpha$ are left as free parameters.
For epochs with two or three frequency data points only, i.e., 2010
October 20, 2010 December 18, 2011 July 30 and 2011 August 01, a systematic
error of 30\% was added to the fit error to account for the low data
statistics. This causes the data points for those epochs to have slightly
larger error bars compared to the epochs with more frequency data
points whose errors are derived directly from the post-fit covariance
matrix. The observed spectral index does not show signs of evolution
and remains steep, i.e. $\alpha$$\sim$$-$1 throughout the period.
This steepness seems to persist persist longer than in the the case
of SN$\,$1993J, whose spectral index evolution shows flattening at
all frequencies beginning at an age $\sim$970 days (\citealp{Alberdi2005},
\citealp{Martividal2011b}, \citealp{Weiler2007})

\begin{table*}
\caption{The SN\,2008iz derived equipartition total minimum energy and magnetic fields.}
\label{table2}
\centering
\begin{tabular}{c c c c c c c c}
\hline
 Date & Days since &  Radius & Luminosity & B$_{eq}$&B$_{eq}$ & E$_{min}$&E$_{min}$\\ 
 (yy/mm/dd) & 18-2-2008 & (cm) & (ergs$^{-1}$) & (G)$_{k = 1}$& (G)$_{k = 2000}$& (ergs)$_{k = 1}$& (ergs)$_{k = 2000}$\\ 
\hline 
2008/03/24 & 36 & 0.88$\times$10$^{16}$ & 142.6$\times$10$^{35}$ & 0.444 & 3.197 & 1.58$\times$10$^{46}$ & 0.82$\times$10$^{48}$\\
2008/05/03 & 76 & 1.67$\times$10$^{16}$ & 126.6$\times$10$^{35}$ & 0.248 & 1.784 & 3.35$\times$10$^{46}$ & 1.74$\times$10$^{48}$\\
2009/04/08 & 416 & 7.22$\times$10$^{16}$ & 13.2$\times$10$^{35}$ & 0.037 & 0.266 & 6.05$\times$10$^{46}$ & 3.14$\times$10$^{48}$\\
2009/04/27 & 435 & 7.51$\times$10$^{16}$ & 8.9$\times$10$^{35}$ & 0.032 & 0.230 & 5.08$\times$10$^{46}$ & 2.63$\times$10$^{48}$\\
2009/09/19 & 580 & 9.61$\times$10$^{16}$ & 6.5$\times$10$^{35}$ & 0.024 & 0.170 & 5.83$\times$10$^{46}$ & 3.02$\times$10$^{48}$\\
2009/10/21 & 612 & 10.07$\times$10$^{16}$ & 4.4$\times$10$^{35}$ & 0.020 & 0.146 & 4.96$\times$10$^{46}$ & 2.57$\times$10$^{48}$\\
2010/10/20 & 976 & 15.04$\times$10$^{16}$ & 3.4$\times$10$^{35}$ & 0.013 & 0.096 & 7.16$\times$10$^{46}$ & 3.71$\times$10$^{48}$\\
2010/12/18 & 1035 & 15.82$\times$10$^{16}$ & 4.0$\times$10$^{35}$ & 0.013 & 0.097 & 8.39$\times$10$^{46}$ & 4.35$\times$10$^{48}$\\
2011/02/12 & 1091 & 16.55$\times$10$^{16}$ & 5.0$\times$10$^{35}$ & 0.014 & 0.099 & 10.10$\times$10$^{46}$ & 5.23$\times$10$^{48}$\\
2011/05/21 & 1189 & 17.82$\times$10$^{16}$ & 4.4$\times$10$^{35}$ & 0.012 & 0.090 & 10.32$\times$10$^{46}$ & 5.35$\times$10$^{48}$\\
2011/08/09 & 1269 & 18.85$\times$10$^{16}$ & 3.3$\times$10$^{35}$ & 0.011 & 0.079 & 9.42$\times$10$^{46}$ & 4.88$\times$10$^{48}$\\
2012/02/02 & 1446 & 21.09$\times$10$^{16}$ & 2.7$\times$10$^{35}$ & 0.009 & 0.068 & 9.70$\times$10$^{46}$ & 5.03$\times$10$^{48}$\\
2012/08/25 & 1651 & 23.64$\times$10$^{16}$ & 4.7$\times$10$^{35}$ & 0.010 & 0.072 & 15.42$\times$10$^{46}$ & 7.99$\times$10$^{48}$\\
2012/11/21 & 1739 & 24.72$\times$10$^{16}$ & 7.1$\times$10$^{35}$ & 0.011 & 0.078 & 20.67$\times$10$^{46}$ & 10.71$\times$10$^{48}$\\
2014/01/23 & 2167 & 29.90$\times$10$^{16}$ & 9.1$\times$10$^{35}$ & 0.010 & 0.071 & 30.23$\times$10$^{46}$ & 15.66$\times$10$^{48}$\\
 \hline 
\end{tabular} 
\tablefoot{
The luminosity is derived from the 22.3\,GHz integrated flux values, while the radius is derived from the VLBI results by assuming a constant expansion index $m$=0.86$\pm$0.02 and c$_y$=(7.5$\pm$0.9)$\times$10$^{-6}$\,arcsec/day.}
\end{table*}

\subsection{Equipartition, total minimum Energy and Magnetic field}
\begin{figure}[h!]
  \resizebox{0.90\hsize}{!}{\includegraphics{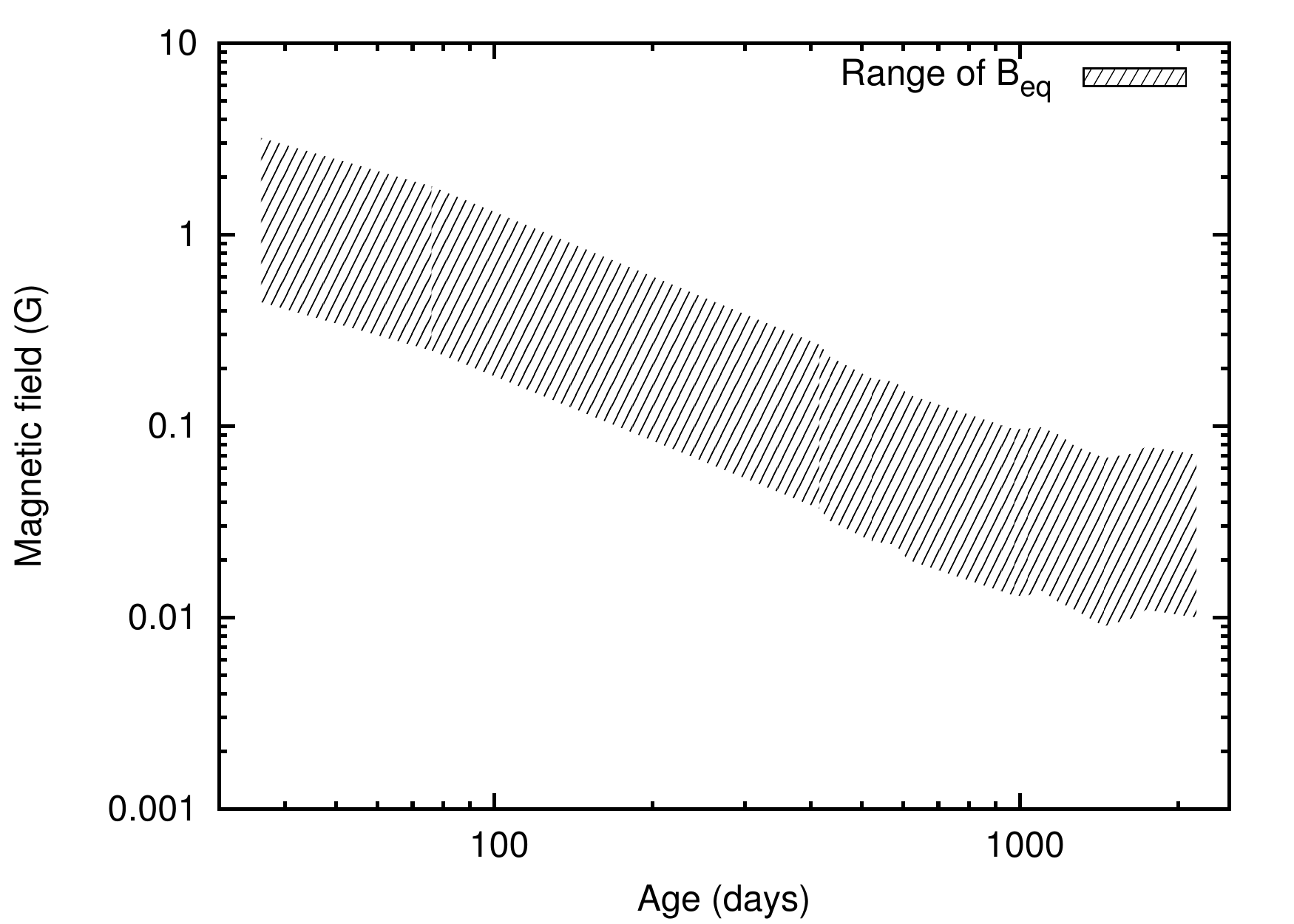}}
  \caption{The 22\,GHz evolution of the magnetic field in the synchrotron emitting plasma of SN\,2008iz with time. The magnetic field evolution is roughly consistent with $B\,\propto\,t^{-1}$ with signs of flattening past day $\sim$\,970.}
 \label{magnetic}
\end{figure}

For sources of radio synchrotron emission, energy equipartition between
the particles and the magnetic field is usually postulated \citep{Pacholczyk1970}.
The total minimum energy content (E$\mathrm{_{min}}$) and magnetic
field (B$_{\mathrm{eq}}$) in SN$\,$2008iz can be expressed as described in relations \ref{eq:E_min} and \ref{eq:B_eq} 

\begin{equation}
E_{min}=c_{13}(1+k)^{4/7}\phi^{3/7}R^{9/7}L^{4/7}\label{eq:E_min}
\end{equation}

and 

\begin{equation}
B_{eq}=4.5^{2/7}c_{12}^{2/7}(1+k)^{2/7}\phi^{-2/7}R^{-6/7}L^{2/7}\label{eq:B_eq}
\end{equation}

\noindent where the dimensionless parameters c$_{12}$ and c$_{13}$ are 9.3$\times$10$^{7}$ and 3.3$\times$10$^{4}$ respectively \citep{Pacholczyk1970} for
spectral index, $\alpha$$\sim$$-$1. The filling factor, $\phi$,
the ratio of the inner and outer radii of the emitting region, is
estimated as 0.3 \citep{Marchili2010}. R is the source radius in
$\mathit{\mathrm{cm}}$ derived from our VLBI data R$_{50\%}$, which
predicts $\mathit{m}$=0.86$\pm$0.02 and c$_{y}$=(7.5$\pm$0.9)$\times$10$^{-6}$$\,$arcsec/day in agreement with previous derivations from \citet{Brunthaler2010}
of $\mathit{m}$=0.89$\pm$0.03. $L$ is the integrated radio luminosity
in ergs$^{-1}$ between the radio synchrotron cutoff frequencies $\nu_{\mathrm{min}}$=10$^{7}$$\,$Hz and $\nu_{\mathrm{max}}$=10$^{11}$$\,$Hz. $\mathit{k}$ is the
ratio between the relativistic heavy particle energy density to the
relativistic electron energy density. The parameter $k$ is dependent
on masses of the particles, and varies from m$_{p}$/m$_{e}$$\approx$1
to m$_{p}$/m$_{e}$$\approx$2$\times$10$^{3}$ \citep{Marchili2010}.
With varying value of $\mathit{k}$, we cannot compute a single value
of total minimum energy and equipartition magnetic field, but can
only give a range of possibilities (see table \ref{table2}).
The derived E$_{min}$ values are in the range of $\sim$10$^{46}$
-- $\sim$10$^{48}$$\,$erg. Our results of B$_{eq}$ at 2008 May
3, which range between 0.25$\,$G and 1.78$\,$G, are comparable to
the values derived by\citet{Marchili2010} at 5$\,$GHz of 0.3$\,$G
and 2.1$\,$G and also on 2009 April 8 between 0.04$\,$G and 0.27$\,$G,
compared to 0.04$\,$G and 0.31$\,$G respectively. Figure \ref{magnetic}
shows the magnetic field evolution as the supernova ages derived from
the integrated 22.3$\,$GHz flux densities from day $\sim$36 to $\sim$2200
after the explosion. From this figure, we notice that the magnetic
fields at each epoch scale with time, showing a hint of flattening
in the time $\geq$ day 1300. This flattening corresponds to the flux-density
flare events when the shock interacts with dense CSM. However, assuming
a constant $k$ value, the magnetic field generally decreases according
to $B\propto\,t^{-1}$ on the full time range.

Considering the equipartition between fields and particles, the B$\mathrm{_{eq}}$
range between (37 -- 266)$\,$mG on 2009 April 8 (day 416) is expected
to explain the observed level of radio emission. The choice of the
radius which is $\sim$7.2$\times$10$^{16}$$\,$cm, allows for a
comparison with other supernovae at about the same point of expansion.
The energy density of the magnetic field in the remnant is postulated
to be lower than the kinetic energy density (i.e., $\mathrm{\rho}_{csm}\mathrm{\upsilon}_{csm}^{2}/2\gtrsim\mathrm{B}_{csm}^{2}/8\pi$),
such that

\begin{equation}
\begin{split}
\mathrm{B}_{csm} & \lesssim\frac{(\dot{\mathrm{M}}\mathrm{\upsilon}_{w})^{1/2}}{\mathrm{r}}\\ & =2.5\left(\frac{\mathrm{\dot{M}}}{10^{-5}\mathrm{M_{\odot}\, yr^{-1}}}\right)^{1/2} \left(\frac{\mathrm{\upsilon}_{w}}{10\,\mathrm{km\, s^{-1}}}\right)^{1/2}\left(\frac{\mathrm{r}}{10^{16}\mathrm{cm}}\right)^{-1}\,\mathrm{mG}.
\label{eq:B_csm}
\end{split}
\end{equation}

\noindent Using a mass-loss of the progenitor of SN$\,$2008iz of $\sim$3.69$\times$10$^{-5}$M$_{\odot}$yr$^{-1}$ \citep{Marchili2010}, a standard pre-supernova wind velocity $\upsilon_{w}$ of 10$\,$km$\,$s$^{-1}$, and a standard CSM density profile of $\mathrm{\rho}_{wind}$$\propto$r$^{-2}$,
we obtain B$_{\mathrm{csm}}$= 0.67$\,$mG, which is a factor of about
55 -- 400 times smaller than the equipartition field. This indicates
that the magnetic field inferred for SN$\,$2008iz cannot originate
solely from compression of the existing circumstellar magnetic field,
which is predicted to increase the field only by a factor of 4 \citep{Dyson1980}.
Large amplification factors of the magnetic field have also been found
for other radio supernova: for SN$\,$2001gd an amplification factor
in the range of 50-350 was determined \citep{Perez-Torres2005}; the
SN$\,$1993J amplification factor of a few hundred \citep{Fransson1998,Perez-Torres2001};
for SN$\,$1986J values in the range of 40--300 \citep{Perez-Torres2002b};
for SN$\,$1979C in the range of 50 -- 400 \citep{Perez-Torres2005};
and between 37 and 260 for SN$\,$2004et \citep{Marti-Vidal2007}.
Thus, if equipartition between fields and particles holds, amplification
mechanisms other than compression of the circumstellar magnetic field
need to be invoked to explain the level of radio emission from the
supernova, such as turbulent amplification due to Rayleigh-Taylor
instability \citep{Chevalier1982,Chevalier1995}.

\newpage
\section{Summary}

We report on multi-frequency VLA and VLBI radio data in the optically thin regime for an on-going monitoring campaign of SN\,2008iz.  We fitted two models to the
data, a simple power-law ($S\,\propto\,t^{\beta}$) and a simplified
Weiler model, yielding $\beta$ = $-$1.22$\pm$0.07 (days 100-1500)
and $-$1.41$\pm$0.02 (days 76-2167) respectively. The light curve
uncovers flux density enhancements at t=970 days and t=1400 days by
a factor of $\sim$2 and $\sim$4 respectively. The later flare, besides
being brighter, does not show signs of decline, at least from the
results examined so far (2014 January 23; day 2167). The flaring activity
is attributed to increase in the number density of shocked circumstellar
medium (CSM) electrons as the expanding shock wave encounters a clumpy
denser medium. The late flare may also be attributed to the transition
of the front shock from the CSM bubble into the interstellar medium
(ISM), although a very dense ISM ($>$10$^{4}$$\,$cm$^{-3}$) would
be required, which is consistent with SN$\,$2008iz being deeply buried
within the center of M$\,$82.

The 1.4$\,$GHz flux density values fall below the supernova model
fit, a phenomena confirmed by 3$\sigma$ LOFAR non-detection limit
at 154$\,$MHz level of 0.41$\,$mJy/beam. The low flux values could
be attributed to free-free absorption (FFA) from a dense foreground
screen along the line of sight, or a low-frequency cut-off caused
by Razin-Tsytovich effect.

The spectrum for SN$\,$2008iz for the period from $\sim$430 to 2167$\,$days
after the supernova explosion shows no signs of evolution and remains
steep ($\alpha$$\thicksim$$-$1). This is different from SN$\,$1993J,
whose spectral index evolution shows $\alpha$ flattening of the spectrum
at all frequencies, beginning at an age $\sim$970$\,$days.

From the 4.8 and 8.4$\,$GHz VLBI images, the supernova expansion
is seen to start with a shell like structure, reflecting an expansion
with similar velocities into all direction that gets more and more
asymmetric at later stages. Finally, in later epochs the structure
breaks up, with bright structures dominating the lower part of the
ring, which serves as an indication of a denser surrounding medium
along the southern direction. This structural evolution differs significantly
from SN$\,$1993J, which remains circularly symmetric over 4000 days
after the explosion. From the size evolution of the SN$\,$2008iz
SNR derived from the 4.8 and 8.4$\,$GHz VLBI images, a deceleration
parameter, $m$, of 0.86$\pm$0.02 and an expansion velocity of (12.1$\pm$0.2)$\times$10$^{3}$$\,$km$\,$s$^{-1}$
are derived for the time range between 73 and 1400 days.

From the energy equipartition between fields and particles, we estimate
the minimum total energy in relativistic particles and the magnetic
fields during the supernova expansion. On 2009 April 8, E$_\mathrm{{min}}$$\propto$3.15$\times$10$^{46}$ -- 1.63$\times$10$^{48}$$\,$erg, which corresponds to an average equipartition magnetic field of B$\mathrm{_{min}}\propto\,$37 -- 266$\,$mG.
We derive the average magnetic field in the circumstellar wind of
B$\mathrm{_{csm}}$= 0.67$\,$mG at a radius from the center of the supernova
explosion of $\sim$7.2$\times$10$^{16}$$\,$cm. Since the supernova
shock compression could only enhance the magnetic field by a factor
of 4, powerful amplification mechanisms must be at play in SN$\,$2008iz
to account for the derived magnetic field amplification of 55 -- 400
that is responsible for the synchrotron radio emission.
\\
\\
We wish to thank Dr.~S.~Mattila for the valuable and constructive comments. Naftali Kimani was supported for this research through a stipend from the International Max Planck Research School (IMPRS) for Astronomy and Astrophysics at the Universities of Bonn and Cologne. This work made use of the Swinburne University of Technology software correlator, developed as part of the Australian Major National Research Facilities Programme and operated under licence. More information on the software can for example be found in \cite{Deller2011}.

\bibliography{sn2008iz.bbl}
\bibliographystyle{aa}

\end{document}